\documentclass{article}
\usepackage[utf8]{inputenc}
\usepackage{amsmath}
\usepackage{hyperref}
\usepackage{tikz,lipsum,lmodern}
\title{Dynamics of tripartite correlations for three qubits in independent thermal reservoirs}
\usepackage{amsfonts}
\usepackage[affil-it]{authblk}
\usepackage[export]{adjustbox}
\usepackage{subcaption}
\usepackage{float}
\usepackage{cite}
\usepackage{setspace}
\usepackage{svg}
\usepackage{xcolor}
\usepackage{graphicx}
\newcommand{\orcid}[1]{\href{https://orcid.org/#1}{\includesvg[width=10pt]{orcid}}}

\usepackage[a4paper,width=150mm,top=35mm,bottom=35mm]{geometry}

\begin{document}
\author{Sourabh Magare, Abhinash Kumar Roy, and Prasanta K. Panigrahi}
\affil{Department of Physical Sciences, Indian Institute of Science Education and Research Kolkata, Mohanpur, West Bengal, 741246, India
}
\affil{\begin{flushleft}\text{\qquad \footnotesize E-mail: \href{mailto:sourabh.magare@gmail.com}{sourabh.magare@gmail.com}, \href{mailto:akr16ms137@iiserkol.ac.in}{akr16ms137@iiserkol.ac.in}, \href{mailto:panigrahi.iiser@gmail.com}{panigrahi.iiser@gmail.com }}  \end{flushleft}}
\date{}
\maketitle
\begin{abstract}
  We investigate the dynamical evolution of multipartite  entanglement in tripartite systems with three qubits present in independent thermal reservoirs, modelled as infinite set of quantum harmonic oscillators. We show that the presence of temperature gradient between reservoirs play a significant role in robustness of multipartite correlation against environmental decoherence. Considering a bilinear form of interaction Hamiltonian, an exact expression for time evolved density matrix is presented. Interestingly, it is observed that the preservation duration of genuine multipartite concurrence for GHZ class Werner state increases as the temperature gradient between reservoirs are increased.  However, this increase is non-linear and the preservation duration is shown to saturate for large temperature gradients. It is also observed that, for large temperature gradient, there are multiple intervals in which coherence and consequently the tripartite negativity shows robustness (freezing of correlation) against the environmental decoherence for W class Werner states. For the common temperature case, we show that the genuine multipartite concurrence for the GHZ class Werner state decays irreversibly, and the state experiences correlation sudden death, with a rate dependent on spectral density of the reservoirs under consideration. For the Ohmic reservoirs at low temperature, we observe that the state retains initial multipartite correlation upto a characteristic time and then sharply decays to zero with a correlation sudden death.  We further investigate the dynamics of tripartite negativity and $l_{1}$ norm of coherence in the W class Werner states, and show irreversible degradation in these correlations. 
\end{abstract}
\section{Introduction}
Quantum entanglement is a fundamental feature of multipartite quantum systems which does not have a classical analog \cite{Horoddecki_RMP_2009,nielsen_chuang}. Apart from playing a crucial role in foundational aspects on quantum theory \cite{Einstein_EPR,schrodinger1936probability,bell1964einstein}, it has found application in various information processing tasks, such as cryptography \cite{Ekert_1991_PRL},  quantum teleportation \cite{bennett1993teleporting}, superdense coding \cite{bennett1992communication_superdensecoding} and sensitive measurements \cite{Pezze2021_nature_enhanced_measurement} to name a few. Quantification for bipartite entanglement is satisfactorily understood through various measures such as  entropy \cite{Vedral_quantifying_ent_PRL_1997}, entanglement of distillation \cite{Bennett_distillation_PRA_1996}, concurrence \cite{hill1997entanglement}, negativity \cite{Vidal_negativity_PRA}, etc. However, a satisfactory extension to quantification of multipartite entanglement is still an open problem \cite{Vidal_J_Mod_opt_2000}.

Multipartite entanglement plays a significant role in illustrating stronger non-locality \cite{GHZ_strong_nonlocality}, and robustness of various information theoretic protocols as compared to the one involving bipartite entanglement \cite{three_party_entanglement1,Hillary_secretsharing}. Furthermore, it has found important applications in the quantum metrology \cite{Giovannetti1330_quantum_meterology_science}, and in the understanding of  quantum-phase transitions\cite{vedral_multipartit_entanglement_RMP}. Therefore, characterization and quantification of multipartite entanglement is of considerable importance and have gained significant attention recently \cite{multiparty_correlation_intresting_complexity_PRL,Ma_PRA_2011,Eberly_2021} . However, even for the simplest case of tripartite qubit systems, a quantification of genuine correlation presents several difficulties \cite{Vidal_J_Mod_opt_2000}. Nevertheless, various measures such as three tangle \cite{wooters_3_tangle_2000}, tripartite negativity \cite{Vidal_negativity_PRA}, genuine multipartite concurrence \cite{Ma_PRA_2011,Eberly_GMC_PRA_2012}, concurrence fill \cite{Eberly_2021}, etc., have been proposed to quantify genuine tripartite entanglement. Out of these measures, genuine multipartite concurrence defined by Rafsanjani et al. \cite{Eberly_GMC_PRA_2012} is a bonafide measure of tripartite entanglement even for mixed states. A closed expression has been found for X type states \cite{Eberly_X_states}, where non-zero terms  in the density matrix are diagonal and anti-diagonal elements only.

For any practical application of entangled systems, interaction of the system with the environment is unavoidable. This interaction of system with environment causes a rapid destruction of the quantum properties of the system. Therefore, it is of considerable importance to study the effects of environment on the dynamics of quantum correlations and possibly device a method to shield quantum system from the environmental interaction to prolong the duration of various quantum properties. In several recent studies, quantum correlations in the presence of a thermal reservoir have been studied, where the reservoir has been modelled as a set of quantized harmonic oscillators \cite{Viola_PRA_1998,Kuang_PRA_199_BEC,Kuang_trapped_Ion_1999,Yuan_2010_J_phys_B, Xu_2011_EPJD, Wu2017_sci_rep}. A single qubit dephasing mechanism has been discussed by Viola and Lloyd \cite{Viola_PRA_1998}. Kuang et al. \cite{Kuang_trapped_Ion_1999} investigated the decoherence in nonclassical motional states of a single trapped ion. For a bipartite system, temperature dependence on the dynamics of correlations has been studied extensively \cite{Xu2010_nature_comm,Yuan_2010_J_phys_B,Eberly_sedden_ent_death_science_2009,decoherence_2010_PRL}. Yuan et al. showed that for a system of two uncoupled qubits, initially prepared in a certain X- type state placed in a common reservoir, quantum correlations can be amplified \cite{Yuan_2010_J_phys_B}. Kuang et al. \cite{Kuang_PRA_199_BEC} studied the dynamics of quantum correlation for two Bose-Einstein condensate, where they investigated the influence of decoherence on quantum coherent atomic tunneling and population difference of two condensates. Recently, dynamics of quantum discord for two initially-correlated qubits in two different ohmic reservoirs has been investigated by Xu et al. \cite{Xu_2011_EPJD}. The quantum correlation for three qubit system has been studied for the case of classical environmental noise, where it is found that the action of different, common or mixed environments has different effects on the robustness of quantum correlations \cite{Tchoffo2016_EPJP_noise}.

In the present work, we study the dynamical evolution of multipartite entanglement in a three qubit systems with each qubit present in an independent thermal reservoir. Considering the thermal reservoirs as a set of infinite quantum harmonic oscillators, and with a bilinear form of non-dissipative interaction Hamiltonian, we provide an exact form of time evolved tripartite density matrix. We show that the temperature gradient between the reservoirs plays a significant role in persistance of multipartite correlations for mixed tripartite systems. Specifically, we study the multipartite correlation measure, genuine multipartite concurrence (GMC) for physically realizable mixed three qubit X state namely, GHZ class Werner state. It is shown that the preservation time for the tripartite correlation decays, and the state experiences the sudden death of GMC for most of the region of mixing parameter, with the decay rate depending on the spectral density of reservoirs under consideration. It is observed that there exist a characteristic time up to which the GMC shows robustness then sharply decays to zero. Remarkably, it is observed that the preservation duration of GMC increases with an increase in temperature gradient between reservoirs. This increase is observed to be non-linear and the preservation duration saturates for very large temperature gradients. We further study the tripartite negativity for W class of Werner state, wherein the preservation duration as affected by the choice of spectral density and the temperature difference among the reservoirs. Interestingly, we observe that for large temperature difference between the reservoirs, there are multiple intervals in which the coherence and the tripartite negativity shows the correlation freezing phenomenon, and hence robustness against the environmental decoherence. 

The paper is organized as follows. In the section \ref{prelim}, we review a few measures of multipartite correlation. In the section \ref{model}, we describe the physical model  of three qubit - thermal reservoir system, and solve for the exact evolution of tripartite density matrix. In the section \ref{dynamics}, using the result obtained we study the dynamical evolution in various tripartite Werner class of states. We conclude after summarizing our results in section \ref{conclusion}.

\section{Tripartite correlation measures}\label{prelim}
\subsection{Genuine multipartite concurrence}
Genuine multipartite concurrence is a well defined measure for genuine multipartite entanglement. It is an entanglement monotone, and becomes zero for all product and bi-separable states, and non-zero for all non biseparable states. For a $N$ qubit density matrix $\rho$, one can have several pure state decomposition of the form,
\begin{equation}
    \rho = \sum_{i}p_{i}^{\lambda}|\psi_{i}^{\lambda}\rangle\langle\psi_{i}^{\lambda}|,
\end{equation}
where, the superscript $\lambda$ refers to specific ensemble. For the above decomposition, the genuine multipartite concurrence is defined as \cite{Ma_PRA_2011},
\begin{equation}
    C_{\lambda}(\rho) = \sum_{i}p_{i}^{\lambda}\left(\min_{j}\big\{\sqrt{2(1-\Pi_{j}(|\psi_{i}^{\lambda}))}\big\}\right),
\end{equation}
where, the index $j$ refers to the possible bipartitions, and $\Pi_{j}$ functional evaluates purity of the subsystems corresponding to the $j$th biparition. Minimum over bipartitions is taken to ensure that, genuine multipartite concurrence must be zero for a biseparable pure state. Finally, the genuine multipartite concurrence for $\rho$ is defined as minimum of $ C_{\lambda}(\rho)$ over all possible pure state decomposition,
\begin{equation}
    C(\rho) = \min_{\lambda} C_{\lambda}(\rho).
\end{equation}
Genuine multipartite concurrence is difficult to evaluate for a general $N$ qubit density matrix, and in fact closed expression only exists for a X type state \cite{Eberly_X_states}, wherein only diagonal elements $\rho_{ii}$ and anti-diagonal elements $\rho_{i(N-i+1)}$ are non-zero, where $i= 1,2,...,N$. For $N$ qubit $X$ state density matrix $\rho^{X}$, the genuine multipartite concurrence is given by \cite{Eberly_2021},
\begin{equation}
    C(\rho^{X}) = 2\max\{0, |\rho_{j(N-j+1)}| - \sum_{k\neq j}^{N/2}\sqrt{\rho_{kk}\rho_{(N-k+1)(N-k+1)}}\},
\end{equation}
where $j = 0,1, ..., N/2$. It is worth noting that the above reduces to the usual concurrence for the two qubit X states.

\subsection{Tripartite negativity}
Consider a three qubit density matrix $\rho_{ABC}$, where the labels $A, B$ and $C$ represent three qubits. There are three bipartitions corresponding to this system namely, $A|BC$, $B|AC$, and $C|AB$. An extension of the usual bipartite negativity \cite{Vidal_negativity_PRA} is defined through geometric mean of the negativity, corresponding to these three bipartitions is a quantifier of tripartite entanglement, known as  tripartite negativity \cite{EPJD_classification_tripartite_negativity}. Corresponding to the bi-partition $A|AB$, the partial transposed density matrix $\rho^{pT}_{A|BC}$ is defined through, $\langle ijk|\rho^{pT}_{A|BC} |lmn\rangle = \langle ljk|\rho_{ABC}|imn\rangle$. The negativity corresponding to the bipartition $A|BC$ is defined as,
\begin{equation}
    \mathcal{N}_{A|BC} = -2\sum_{i}\lambda_{i}^{A|BC},
\end{equation}
where, $\lambda_{i}^{A|BC}$ are the negative eigenvalues of the partially transposed density matrix $\rho^{pT}_{A|BC}$. Similarly one can find the negativity $\mathcal{N}_{B|AC}$ and $\mathcal{N}_{C|AB}$, corresponding to the bipartitions $B|AC$ and $C|AB$, respectively. The tripartite negativity is defined as \cite{EPJD_classification_tripartite_negativity},
\begin{equation}
    \mathcal{N}(\rho_{ABC}) = \left( \mathcal{N}_{A|BC}~\mathcal{N}_{B|AC}~\mathcal{N}_{C|AB}\right)^{\frac{1}{3}},
\end{equation}
which is the geometric mean of the negativity corresponding to all three bipartitions.
\section{Model: three qubits in independent thermal reservoirs}\label{model}

We consider the total system as three non-interacting qubits labeled as A, B, C and three independent heat baths, governed by the Hamiltonian 
\begin{equation}\label{totalHamiltonian}
    \mathcal{H}_{T} = \mathcal{H}_{S} + \mathcal{H}_{B} + \mathcal{H}_{I},   
\end{equation}
where, $\mathcal{H}_{S}$ is the Hamiltonian of three qubits, $\mathcal{H}_{B}$ is the Hamiltonian of the heat baths and $\mathcal{H}_{I}$ is the Hamiltonian representing interaction between three qubits and their respective reservoirs. Hamiltonian of the three qubits is given by,
\begin{equation}
    \mathcal{H}_{S} = \frac{\Omega_{A}}{2}\sigma_{z}^{A} + \frac{\Omega_{B}}{2}\sigma_{z}^{B} + \frac{\Omega_{C}}{2}\sigma_{z}^{C},
\end{equation}
where, $\sigma_{z}^{i} = |0\rangle_{i}\langle 0| + |1\rangle_{i}\langle 1|$ with $i = A, B, C$. States $|0\rangle_{i}$ and $|1\rangle_{i}$ are the excited and ground state respectively, of the $i^{th}$ qubit. Each heat bath is modelled by an infinite set of harmonic oscillators. The Hamiltonian of the reservoirs is given by,
\begin{equation}
    \mathcal{H}_{R} = \sum_{k}\omega_{a k}a_{k}^{\dagger}a_{k}+\omega_{b k}b_{k}^{\dagger}b_{k}+\omega_{c k}c_{k}^{\dagger}c_{k},
\end{equation}
where, $\omega_{i k}$ is the frequency of $k$th harmonic oscillator of $i$th qubit. $a_{k}^{\dagger}$, $b_{k}^{\dagger}$ and $c_{k}^{\dagger}$ are the creation operators and $a_{k}$, $b_{k}$ and $c_{k}$ are the annihilation operators for an oscillator in $k$th mode. The interaction Hamiltonian of the system and bath is assumed to be of the following form,
\begin{equation}
    \mathcal{H}_{I} = \sum_{k} (\sigma_{z}^{A}\Omega_{A}g_{A k}a_{k} + \sigma_{z}^{B}\Omega_{B}g_{B k}b_{k} + \sigma_{z}^{C}\Omega_{C}g_{C k}c_{k} + h.c.),
\end{equation}
where $g_{i k}$ is the interaction strength of $i$th qubit with the $k$th mode of harmonic oscillator.

\begin{figure}[ht!]
    \centering
    \includegraphics[width=0.75\textwidth]{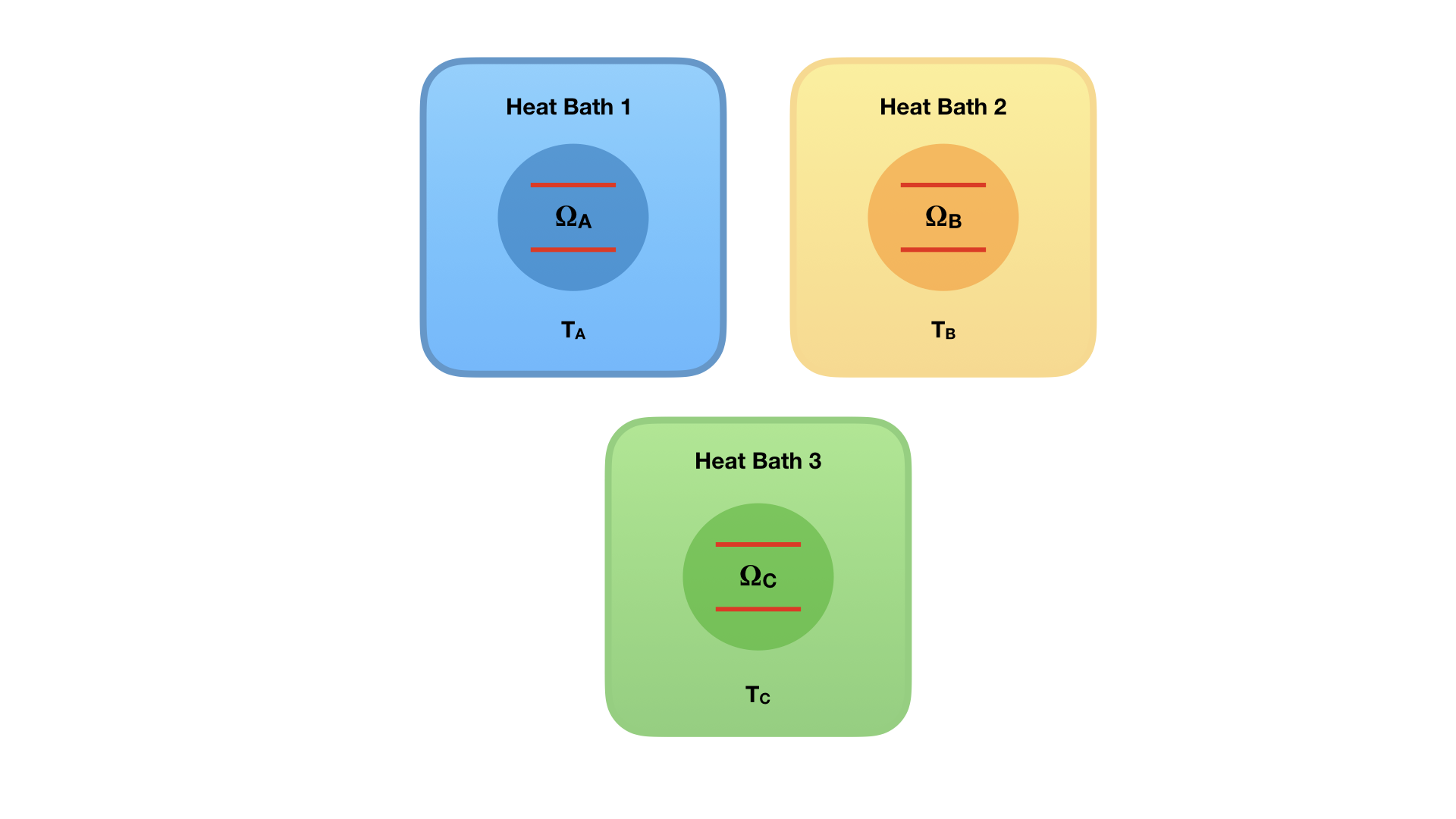}
    \caption{Three non-interacting qubits $A, B$ and $C$ in three independent environments with temperatures $T_{A}$, $T_{B}$ and $T_{C}$, respectively.}
    \label{area}
\end{figure}

As $[\mathcal{H}_{I},\mathcal{H}_{S}] = 0$, the system $\mathcal{H}_{S}$ is conservative and therefore represents a purely decohering mechanism.
It is convenient to solve the problem in the interaction picture associated to $\mathcal{H}_{S} + \mathcal{H}_{R}$, the effective Hamiltonian then becomes,
\begin{equation}
    \mathcal{\Tilde{H}}(t) = \sum_{k} (\sigma_{z}^{A} e^{-i\omega_{a k}t}\Omega_{A}g_{A k}a_{k} + \sigma_{z}^{B}e^{-i\omega_{b k}t}\Omega_{B}g_{B k}b_{k} + \sigma_{z}^{C}e^{-i\omega_{c k}t}\Omega_{C}g_{C k}c_{k} + h.c.).
\end{equation}
It is to be noted that this form of Hamiltonian has been studied earlier by
Xu et. al. \cite{Xu_2011_EPJD} for the case of bipartite system and Viola et. al \cite{Viola_PRA_1998} for the case of single qubit.
The time evolution operator generated by the effective Hamiltonian $\mathcal{\Tilde{H}}(t)$ is given by $U = U_{a}(t)\otimes U_{b}(t)\otimes U_{c}(t)$, where,

\begin{equation}\label{unitary}
    U_{i} = exp\Bigg\{ \frac{\sigma_{z}^{i}}{2}\sum_{k}(X^{\dagger}_{k}\xi_{X k}(t) - h.c.) \Bigg\}.
\end{equation}
Here $\xi_{X k} = \frac{2g_{X k}\Omega_{X}}{w_{X k}}(1-e^{i w_{X k}t})$ and $X = a,b,c$ when $i = A, B, C$ respectively.
To study the dynamics of correlation of this tripartite system, we need to find the time evolution of the density matrix $\Tilde{\rho}(t)$. In this case it becomes possible to find $\Tilde{\rho}(t)$ without any approximations.
We assume that the initial state of the three heat baths (environment) is given by the density operator:
\begin{equation}
    \Tilde{\rho}_{E}(0) = \Pi_{X, k}(1-e^{-\beta_{X}\omega_{X k}})e^{-\beta_{X}\omega_{X k}X_{k}^{\dagger}X_{k}}.
\end{equation}
Here $\beta_{X} = (k_{B}T_{X})^{-1}$ and $\Tilde{\rho}_{E}(0)$ can be written as the tensor product of each subsystem $\Tilde{\rho}_{E}(0) = \Tilde{\rho}_{E A}\otimes\Tilde{\rho}_{E B}\otimes\Tilde{\rho}_{E C}$. The inital state of the three qubits is denoted by the density operator $\Tilde{\rho}_{S}(0)$, the total system is then given by tensor product of the system and the environment states $\Tilde{\rho}(0) =\Tilde{\rho}_{S}(0)\otimes \Tilde{\rho}_{E}(0)$. The reduced density matrix of the system at any time $t$ is obtained by taking a trace over the environment,
\begin{equation}
    \Tilde{\rho}_{S}(t) = Tr_{E}[U^{\dagger}(t)\Tilde{\rho}(0) U(t)].
\end{equation}

The set $\{|mnl\rangle\}$, where $m,n,l = 0,1$, forms a basis of the three qubits system. The energy eigenvalue of $\mathcal{H}_{S}$ in these basis is given by $E_{mnl} = (-1)^{m}\Omega_{A} + (-1)^{n}\Omega_{B} + (-1)^{l}\Omega_{C}$.

Then, the density matrix of the system evaluates to 
\begin{equation}
    \Tilde{\rho}_{(mnl)(m'n'l')}(t) = \Tilde{\rho}_{(mnl)(m'n'l')}(0)F_{(mnl)(m'n'l')}(t).
\end{equation}
The factor $ F_{(mnl)(m'n'l')}(t)$ depends only on the reservoir and is given by,
\begin{equation}
\begin{aligned}
    F_{(mnl)(m'n'l')}(t) = \Pi_{k}Tr_E[D_{A}(-\alpha_{m'n'l'})D_{A}(\alpha_{mnl})\Tilde{\rho}_{a k}]&Tr_E[D_{B}(-\beta_{m'n'l'})D_{B}(\beta_{mnl})\Tilde{\rho}_{b k}]\\&\times Tr_E[D_{C}(-\gamma_{m'n'l'})D_{C}(\gamma_{mnl})\Tilde{\rho}_{c k}].
    \end{aligned}
\end{equation}

Here, $\alpha_{mnlk} = E^{a}_{mnl}\xi_{ak}(t)$, $\beta_{mnlk} = E^{b}_{mnl}\xi_{bk}(t)$, $\gamma_{mnlk} = E^{c}_{mnl}\xi_{ck}(t)$ and $D_{i}(\alpha) = \operatorname{exp}(\alpha a^{\dagger}_{i} - \alpha^{*}a_{i})$ is the displacement operator with subscript $i$ representing the $i$th reservoir and $a_{i}(a_{i}^{\dagger})$ being annihilation (creation) operators for the corresponding reservoir. Using the property of the displacement operator, $ D(\alpha_{1})D(\alpha_{2}) = D(\alpha_{1} + \alpha_{2})\exp{(i~\operatorname{Im} (\alpha_{1}\alpha_{2}))}$ and the result \cite{Kuang_trapped_Ion_1999}, $\operatorname{Tr}_{E}[D(\alpha)\Tilde{\rho}_{k}] = \operatorname{exp}\left(-{\frac{1}{2}|\alpha|^{2}\coth\frac{\beta \omega_{j}}{2}}\right)$ 
 we obtain the expression for $F_{(mnl)(m'n'l')}(t)$ as follows,
\begin{equation}\label{F1}
\begin{aligned}
    F_{(mnl)(m'n'l')}(t) = \operatorname{exp}\left[(\delta_{m,m'} - 1)\Gamma_{A}(t)\right]&\operatorname{exp}\left[(\delta_{n,n'} - 1)\Gamma_{B}(t)\right]\\& \times\operatorname{exp}\left[(\delta_{l,l'} - 1)\Gamma_{C}(t)\right],
    \end{aligned}
\end{equation}
where $m,n,l = 0,1$ , the real functions,
\begin{equation}
    \Gamma_{X}(t) = \sum_{k} \frac{8 \Omega^{2}_{X}|g_{X,k}|^{2}}{\omega^{2}_{X k}} \coth\left({\frac{\beta_{X}\omega_{Xk}}{2}}\right) \sin^{2}\left({\frac{\omega_{X}t}{2}}\right),
\end{equation}
entirely describes the decoherence mechanism of the tripartite system, and $\delta_{i,j}$ is the Kronecker delta.
Assuming that the states of the reservoir are very dense and taking a continuum limit changes summation over $k$ to an integral as
 $\sum_{k}\rightarrow \int d\omega_{X} J(\omega_{X})$, where $J(\omega_{X}) $is the spectral density which characterises the reservoir. With $J(\omega_{X}) = \delta(\omega_{X} - \omega_{X,k})|g_{X, k}|^{2}$ we can rewrite $\Gamma_{X}(t)$ as,
\begin{equation}\label{dec1}
    \Gamma_{X}(t) = 8\int_{0}^{\infty}d\omega_{X}\frac{J(\omega_{X})}{\omega_{X}^{2}}\Omega^{2}_{X}\coth\left({\frac{\beta_{X}\omega_{X}}{2}}\right) \sin^{2}\left({\frac{\omega_{X}t}{2}}\right).
\end{equation}

We now take recourse to the Schr{\"o}dinger picture. The time evolved density matrix inherits an oscillatory term and takes the following form,

\begin{equation}\label{rho1}
    \rho_{(mnl)(m'n'l')}(t) = \rho_{(mnl)(m'n'l')}(0)F_{(mnl)(m'n'l')}(t)e^{-i(E_{mnl} - E_{m'n'l'})t}.
\end{equation}
With this exact time evolved tripartite density matrix, we are now well equipped to study the dynamics of multipartite correlation. As evident from the above expression, the diagonal elements will remain invariant, however, the off-diagonal elements inherit a phase along with a damping factor, dependent of the spectral density, coupling constant, and the temperature of the reservoirs. We note in passing that, appearance of hyperbolic cotangent term in the decaying factor plays a crucial role in the environmental interaction yielding a noise term in the classical limit, i.e., $\hbar \rightarrow 0$ limit. For certain choices of spectral density, the integrand in the decaying factor (\ref{dec1}) will have the form $\sim \hbar\omega \cos(\omega t)\coth(\hbar \beta\omega/2)$ that upon integration leads to a derivative in $\coth (\pi t/\beta\hbar)$, which using the formula $\coth'(x) = -\operatorname{csch}^{2}(x) + 2\delta(x)$ results in a delta function term (noise term) in the classical limit $\hbar \rightarrow 0$ \cite{Ford1996_Nature}.

In the following section, we use the obtained time evolved density matrix to study the evolution of genuine tripartite correlations in various initially correlated three qubit density systems.

\section{Dynamics of tripartite correlations}\label{dynamics}
\subsection{Evolution of GMC in GHZ class Werner state }
For a system of three non-interacting qubits in three independent thermal reservoir we consider the initial state to be $\rho(0) = \rho^{W}$, where $\rho^{W}$ is a three qubit Werner state given by
\begin{equation}\label{werner}
\rho^{W} = x|\psi\rangle\langle\psi| + (1-x)\frac{I}{8}, 
\end{equation}
where, $I$ is the identity matrix, $x\in [0,1]$ is the mixing parameter, and $|\psi\rangle$ a three qubit pure state. As evident from the above expression, Werner state is a convex sum of maximally mixed state and a pure state. In the following, we study the evolution of multipartite correlation for various choice of tripartite entangled pure state $|\psi\rangle$ in $\rho^{W}$. 

We consider first the maximally entangled GHZ state, $|GHZ\rangle = \frac{1}{\sqrt{2}} (|000\rangle + |111\rangle)$ as the pure state in the Werner state. With this choice the initial density matrix becomes,
\begin{equation}\label{wghz}
    \rho(0) = x|GHZ\rangle\langle GHZ| + (1-x)\frac{I}{8}.
\end{equation}
As evident from the explicit expression for time evolved density matrix (\ref{F1}) and (\ref{rho1}), the diagonal elements will remain unchanged, and only the off-diagonal element gains a phase factor and decaying factor. The only non zero off-diagonal elements for (\ref{wghz}), are the the terms $\rho_{(000)(111)}$ and $\rho_{(111)(000)}$. Therefore, the time evolution of this off-diagonal term is given by,
\begin{equation}\label{den1}
    \rho_{(000)(111)}(t) = \rho_{(000)(111)}(0)F_{(000)(111)}(t)e^{-i(E_{000} - E_{111})t}.
\end{equation}
Using (\ref{F1}), (\ref{den1}), we obtain, 
\begin{equation}
    \rho_{(000)(111)}(t) = \frac{x}{2}e^{-2i(\Omega_{A} + \Omega_{B} + \Omega_{C})t}e^{-(\Gamma_{A}(t) +\Gamma_{B}(t) + \Gamma_{C}(t))},
\end{equation}
where $\Omega_{A}, \Omega_{B}$ and $\Omega_{C}$ are the characteristic of three qubits $A, B$ and $C$ respectively, and $\Gamma_{A}, \Gamma_{B}$ and $\Gamma_{C}$ are the decaying factors associated with the respective reservoirs,, where qubits are placed, depends on the spectral density and temperature of the reservoir.  The genuine multipartite concurrence for $\rho(t)$ is obtained as,
\begin{equation}\label{a1}
    C_{GM}(\rho(t)) = \operatorname{max}\{0, x e^{-(\Gamma_{A}(t) +\Gamma_{B}(t) + \Gamma_{C}(t))} - \frac{3}{4}(1-x) \}.
\end{equation}
We now have the following situation: For a choice of spectral density $J(\omega_{X})$, we can evaluate the decohering function $\Gamma_{X}(t)$ for each reservoir where $X = \{A,B,C\}$ and study the dynamics of genuine multipartite concurrence from the expression (\ref{a1}). A few observations can be qualitatively stated. Observing that the decohering function  $\Gamma_{X}(t)$ is a real, positive and increasing function of time, it is evident that genuine multipartite concurrence decreases with time for a given mixing parameter. Furthermore, the domain of mixing parameter for which the GHZ class Werner state is tripartite entangled, shrinks with increasing time. Since $\Gamma_{X}(t)$ also depends on the spectral density $J(\omega_{X})$ and $\Omega_{X}$ of the system, the rate of decrease of genuine multipartite concurrence is characterised by the type of reservoir, the system and the interaction parameter of the reservoir with the system.

\begin{figure}[h]
\begin{subfigure}{0.5\textwidth}
\centering
\includegraphics[width=0.85\textwidth, height = 5cm]{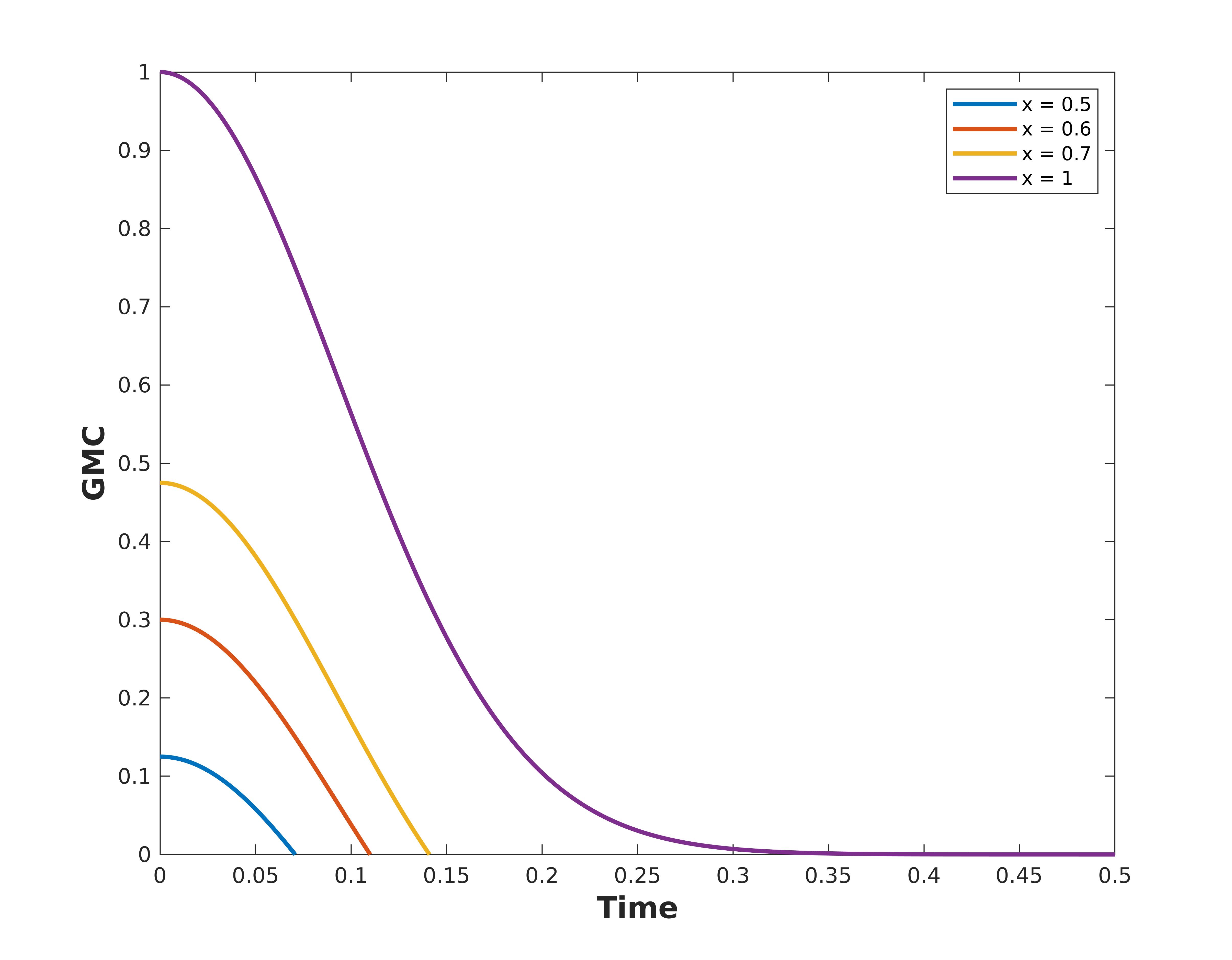} 
\caption{}
\end{subfigure}
\begin{subfigure}{0.5\textwidth}
\centering
\includegraphics[width=0.85\textwidth, height = 5cm]{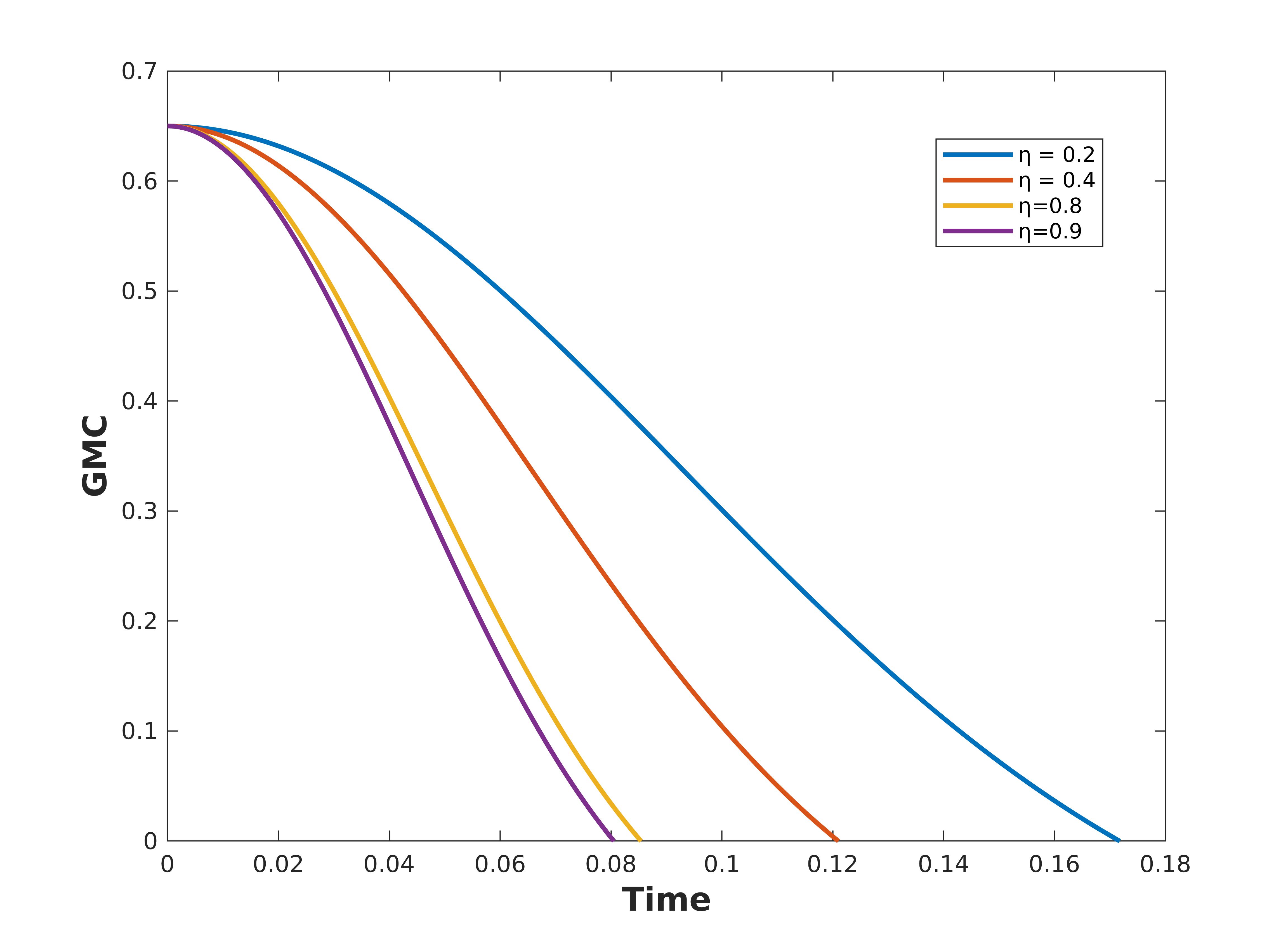}
\caption{}
\end{subfigure}
\caption{Genuine multipartite concurrence (GMC) plotted against time (in the units of $\omega_{c}$) for Ohmic spectral density at zero temperature (a) for various mixing parameter at fixed $\eta = 0.2$, $\Omega^{2} = 12$ (b) for various interaction parameter $\eta$ at a fixed mixing parameter $x = 0.8$, $\Omega = 12$. Entanglement sudden death is observed for $x<1$.}
\label{GMC_ohmic_zero_T}
\end{figure}

Considering each reservoir to be characterized by the ohmic spectral density, given by \cite{Leggett_review},
\begin{equation}
    J(\omega_{A}) = J(\omega_{B}) = J(\omega_{C}) = J(\omega) = \eta\omega e^{-\omega / \omega_{c}},
\end{equation}
where $\omega_{c}$ is the cutoff frequency and $\eta$ is the system-reservoir coupling constant.
It is straightforward to evaluate the decohering function $\Gamma_{X}(t)$  for reservoir with ohmic spectral density at zero temperature and one obtains,
\begin{equation}
    \Gamma_{X}(t) = 2\Omega_{X}^{2}\eta \operatorname{ln}(1+ (\omega_{c}t)^{2}).
\end{equation}
The genuine multipartite concurrence is obtained as,
\begin{equation}\label{28}
    C_{GM}(\rho(t)) = \operatorname{max}\{0, x (1+(\omega_{c}t)^{2})^{-2\eta \Omega^{2}}  - \frac{3}{4}(1-x) \},
\end{equation}
with $\Omega^{2} = \Omega_{A}^{2} + \Omega_{B}^{2} + \Omega_{C}^{2}$.
We observe that genuine multipartite concurrence falls more rapidly as the value of $\eta$ is increased. Thus, more strongly the system is coupled to the reservoir, faster is the degradation in correlations of the system. Moreover, if the initial value of mixing parameter $x$ is larger, correlations of the system last for a longer time.
The preservation time $t_{p}$ is the time interval for which the correlations remain non-zero. For the case of ohmic spectral density at zero temperature, the preservation time is calculated to be
\begin{equation}\label{29}
    t_{p} = \frac{1}{\omega_{c}}\sqrt{\left(\frac{4x}{3(1-x)}\right)^{1/2\eta \Omega^{2}} - 1}.
\end{equation}
From Fig (\ref{GMC_ohmic_zero_T}), we infer that greater the initial value of the mixing parameter $x$, the larger will be the preservation time. It is to be noted that for the state with mixing parameter $x = 1$, the preservation time is infinite. Moreover, the preservation time decreases rapidly as the coupling constant of the system is increased. Consequently, preservation time for the system which is strongly coupled to the reservoir will be smaller as compared to the system which is loosely coupled to the reservoir. It is interesting to note that the preservation time also depends on the types of two level system under consideration. When the energy gap between the two levels is larger, i.e., for larger $\Omega$, and since $\Omega$ plays the same role as $\eta$ in the expression (\ref{28}) and (\ref{29}), it is evident that the preservation time is reduced. Therefore,  a three qubit system for which correlation lasts for a significant duration, one has to decrease the coupling constant $\eta$, the energy gap $\Omega$ and increase the initial mixing parameter $x$.

In general, it is difficult to evaluate $\Gamma_{X}$ for the case  of ohmic reservoir at finite temperature. However, it is possible to find a closed expression in the limit of low temperature. 
Let $T_{A},T_{B}$ and $T_{C}$ be the temperatures of reservoirs $A, B, C$ respectively. 
For the case of finite but small temperatures of the ohmic reservoirs, the decohering function is calculated to be\newline
$\Gamma_{X} =  2\Omega_{X}^{2}\eta \operatorname{ln}\left[\left(1+ (\omega_{c}t)^{2}\right)\frac{\beta_{X}^{2}}{\pi^{2}t^{2}}\sinh^{2}{\left(\frac{\pi t}{\beta_{X}}\right)}\right]$. One then obtains the genuine multipartite concurrence as, 
\begin{equation}
    C_{GM}(\rho(t)) = \operatorname{max}\{0, x \left(\frac{(1+(\omega_{c}t)^{2})}{\pi^{2}t^{2}}\beta^{2}_{A}\beta^{2}_{B}\beta^{2}_{C}\sinh^{2}{\left(\frac{\pi t}{\beta_{A}}\right)}\sinh^{2}{\left(\frac{\pi t}{\beta_{B}}\right)}\sinh^{2}{\left(\frac{\pi t}{\beta_{C}}\right)}\right)^{-2\eta \Omega^{2}}  - \frac{3}{4}(1-x) \}
\end{equation}

\begin{figure}[h]
    \centering
    \includegraphics[width=0.5\textwidth]{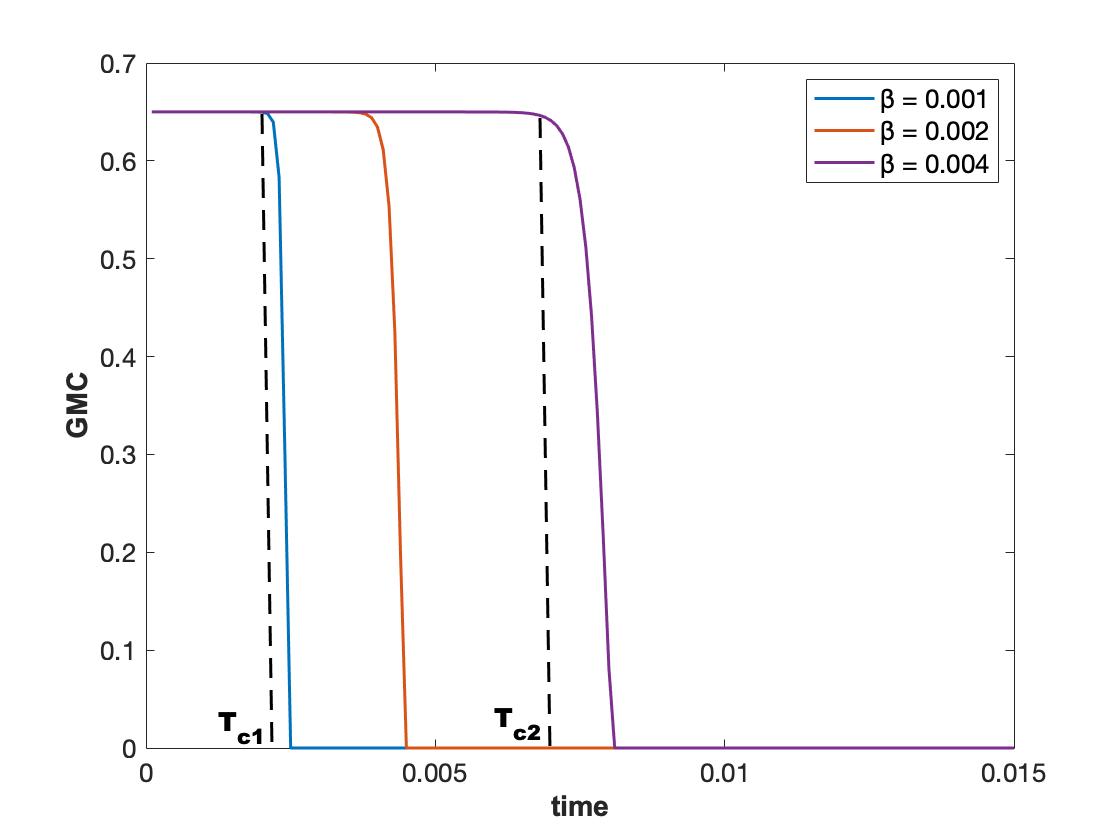}
    \caption{Genuine multipartite concurrence (GMC) as a function of time for different values of $\beta$ for $x = 0.8$, $\eta = 0.4$ and $\Omega^{2} = 36 $ along with $\beta_{A} = \beta_{B} = \beta_{C}$ = $\beta$.} \label{ptime1}
\end{figure}
    
\begin{figure}[ht]
    \centering
    \includegraphics[width=0.5\textwidth]{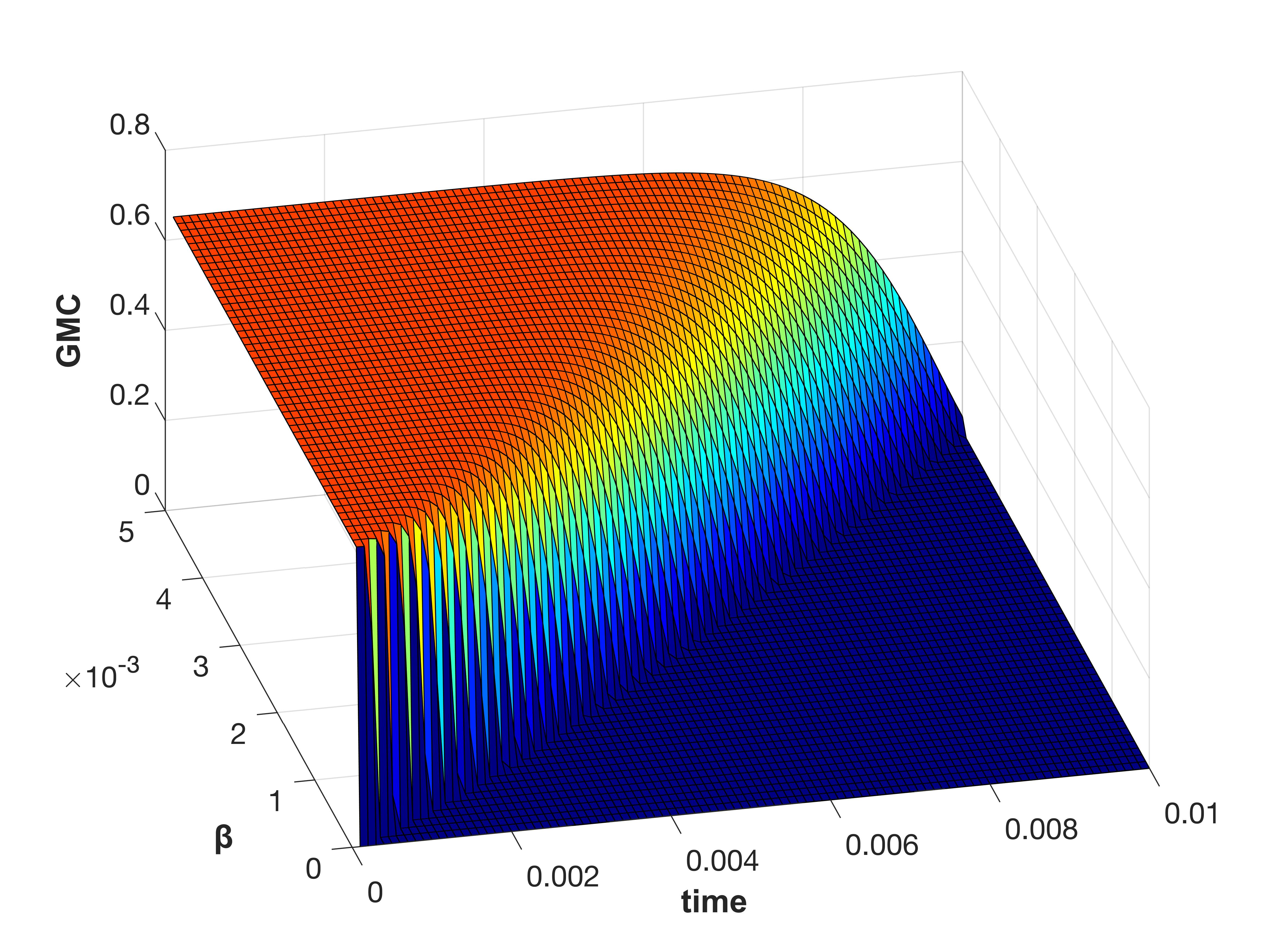}
    \caption{ 3D plot of Genuine multipartite concurrence (GMC) as a function of inverse temperature $\beta$ and time for Ohmic reservoir at a fixed mixing paramter $x = 0.8$, $\eta = 0.4$ and $\Omega^{2} = 36 $ along with $\beta_{A} = \beta_{B} = \beta_{C}$ = $\beta$. For the time interval $0< t < T_{c}$, the state retains initial correlations then sharply decays to zero. As temperature is increased, correlations decay faster.}
    \label{3dplot1}
\end{figure}

 In the Figure (\ref{3dplot1}), genuine multipartite concurrence (GMC) is plotted against the inverse of temperature ($\beta$) and time ($t$) for the mixing parameter $x = 0.8$, $\Omega^{2} = 36$, $\eta = 0.4$ and $\beta_{A} = \beta_{B} = \beta_{C}$ = $\beta$ i.e all three reservoirs are at same temperature. It is observed that the correlations decay at all temperatures. Interestingly, we observe that in the time interval $0 < t < T_{c}$, the correlations retain their initial value and then decay rapidly to zero. Characteristic time $T_{c}$ is the time interval for which the correlations does not decay from their initial value. From Fig. (\ref{ptime1}), we observe that the characteristic time $T_{c}$ increases with decrease in temperature i.e.,  smaller the value of temperature (larger value of $\beta$), longer is the retention of the correlations. After this retention period, the state experiences correlation sudden death. We observe that as we increase the value of the temperature, the state suffers rapid decay in correlations. Thus, smaller temperatures have longer preservation time. 

We can also find the preservation time $t_{p}$ as an implicit function of temperature of the three reservoirs,
\begin{equation}
    \left(\beta_{A}\beta_{B}\beta_{C}\sinh{\left(\frac{\pi t_{p}}{\beta_{A}}\right)}\sinh{\left(\frac{\pi t_{p}}{\beta_{B}}\right)}\sinh{\left(\frac{\pi t_{p}}{\beta_{C}}\right)}\right)^{2} = \frac{\pi^{2} t_{p}^{2}}{(1 + (\omega^{2}_{c} t_{p}^{2}))}\left( \frac{4x}{3(1-x)}\right)^{1/2\eta\Omega^{2}}.
\end{equation}
To discuss the case of different temperatures of three reservoirs on the correlations and preservation time,  it is convenient to consider  $\beta_{B} = k_{1}  \beta_{A}$ and $\beta_{C} = k_{2} \beta_{A}$, where $k_{1}$ and $k_{2}$ are the factors by which $\beta_{B}$ and $\beta_{C}$ differ from $\beta_{A}$. In the following, we analyse the dynamics of preservation time of genuine multipartite concurrence against $\beta_{A}$ for different values of $k_{1}$ and $k_{2}$. 
\begin{figure*}[h!]
   \subfloat[\label{}]{%
      \includegraphics[trim=20 40 50 30,clip, width=0.3\textwidth]{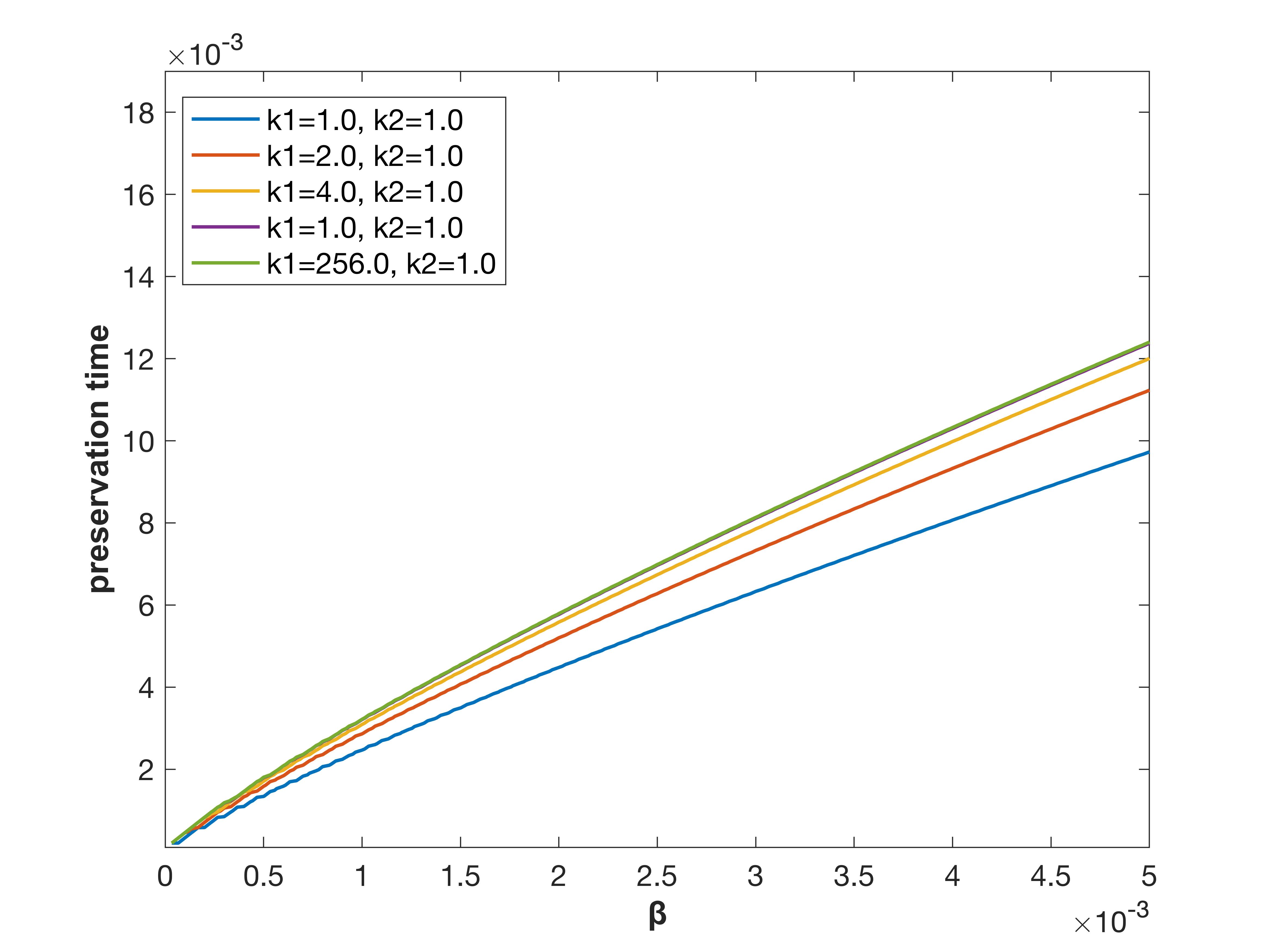}}
\hspace{\fill}
   \subfloat[\label{} ]{%
      \includegraphics[trim=20 40 50 30,clip, width=0.3\textwidth]{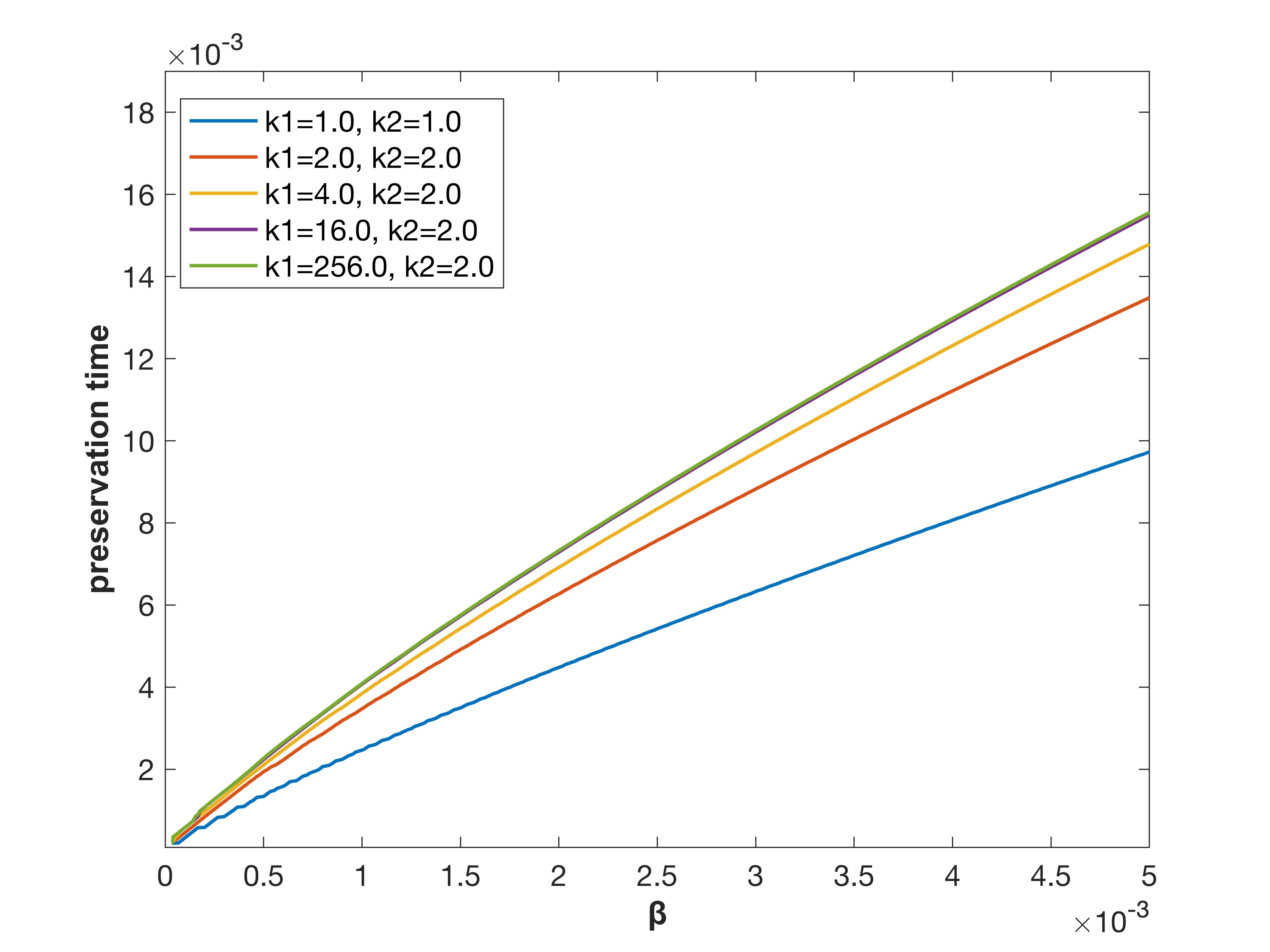}}
\hspace{\fill}
   \subfloat[\label{}]{%
      \includegraphics[trim=20 40 50 30,clip, width=0.3\textwidth]{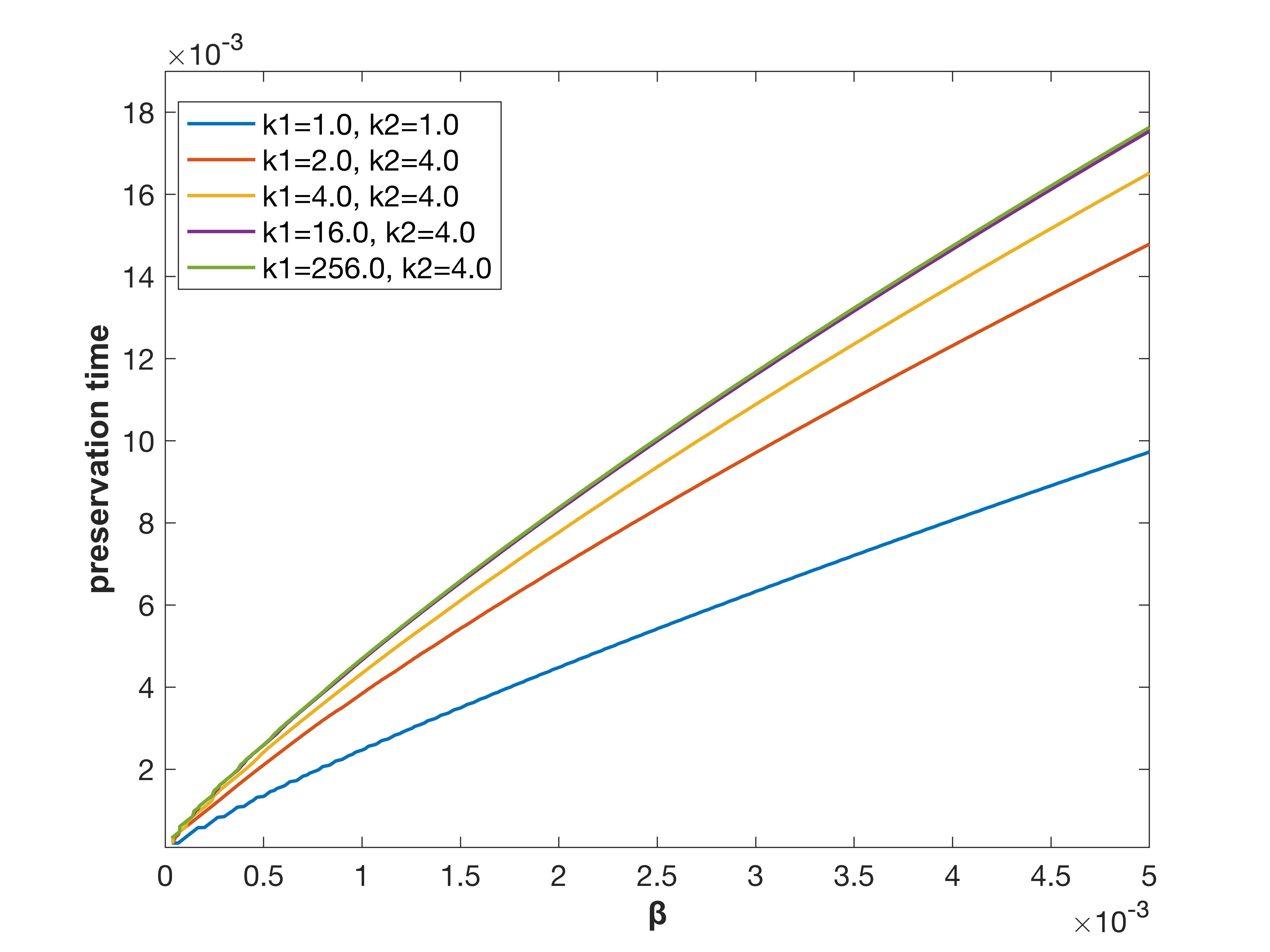}}\\
      \centering
   \subfloat[\label{}]{%
      \includegraphics[trim=20 40 50 30,clip, width=0.3\textwidth]{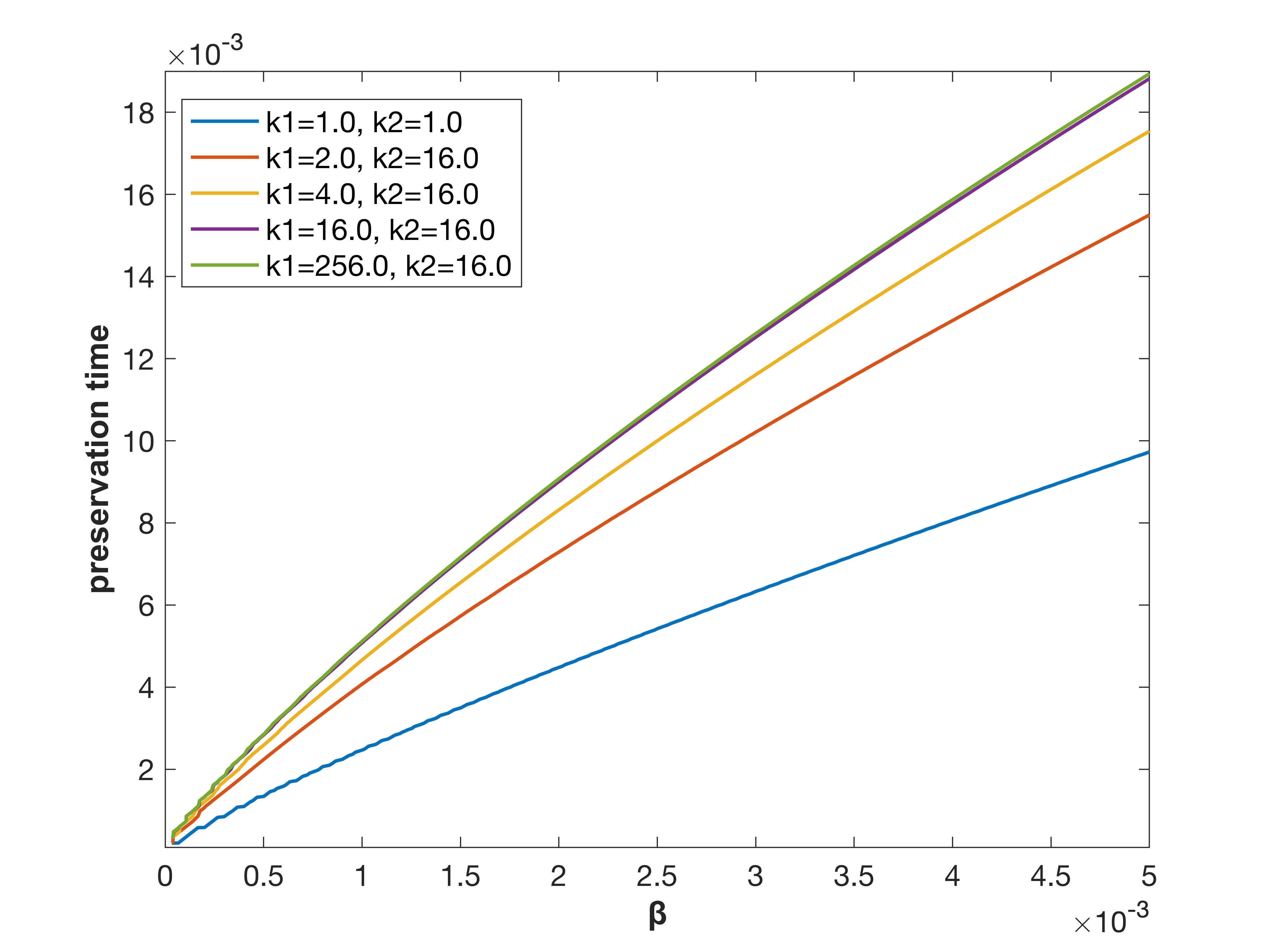}}
   \subfloat[\label{}]{%
      \includegraphics[trim=20 40 50 30,clip, width=0.3\textwidth]{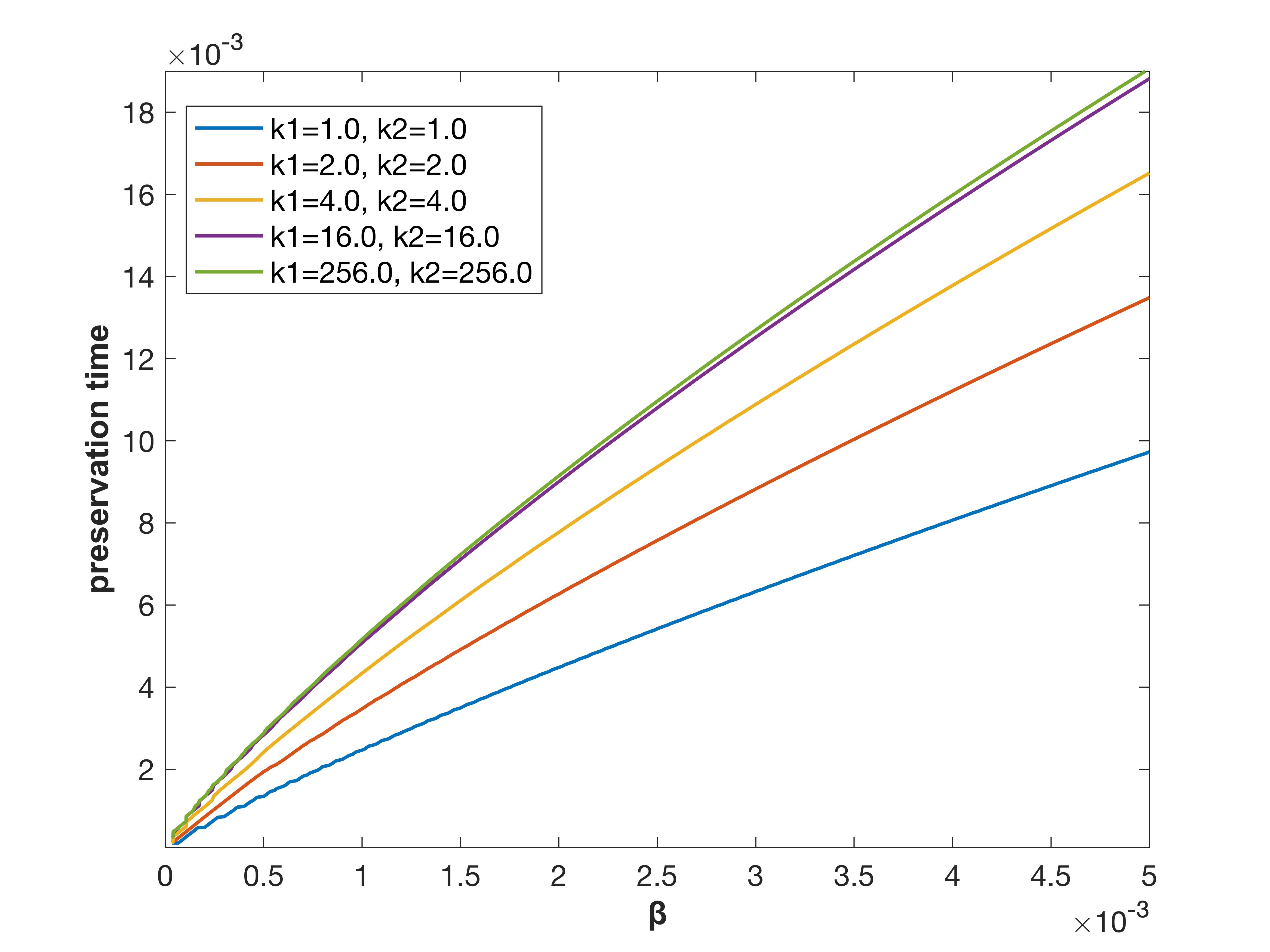}}
\caption{ Preservation time is plotted as the function of inverse of temperature ($\beta$) for different values of $k_{1}$ and $k_{2}$ and for the parameters $x = 0.8$, $\eta = 0.2$ and $\Omega^{2} = 36$ where $\beta_{A} = \beta $, $\beta_{B} = k_{1}  \beta$ and $\beta_{C} = k_{2} \beta$. }\label{GMC_ohmic_differnt_temperatures}
\end{figure*}

\begin{figure*}[h!]
   \subfloat[\label{}]{%
      \includegraphics[trim=20 40 50 30,clip, width=0.3\textwidth]{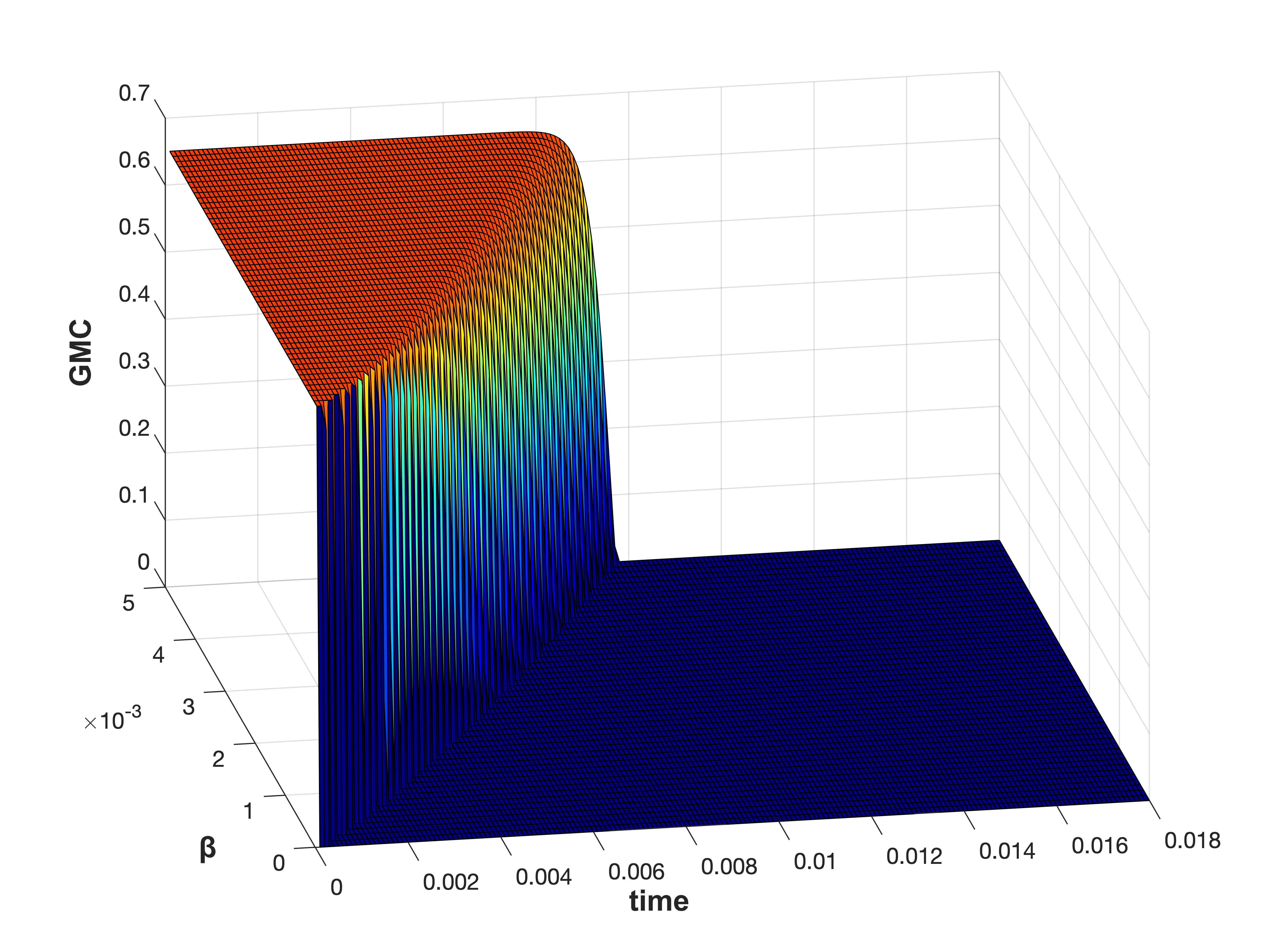}}
\hspace{\fill}
   \subfloat[\label{} ]{%
      \includegraphics[trim=20 40 50 30,clip, width=0.3\textwidth]{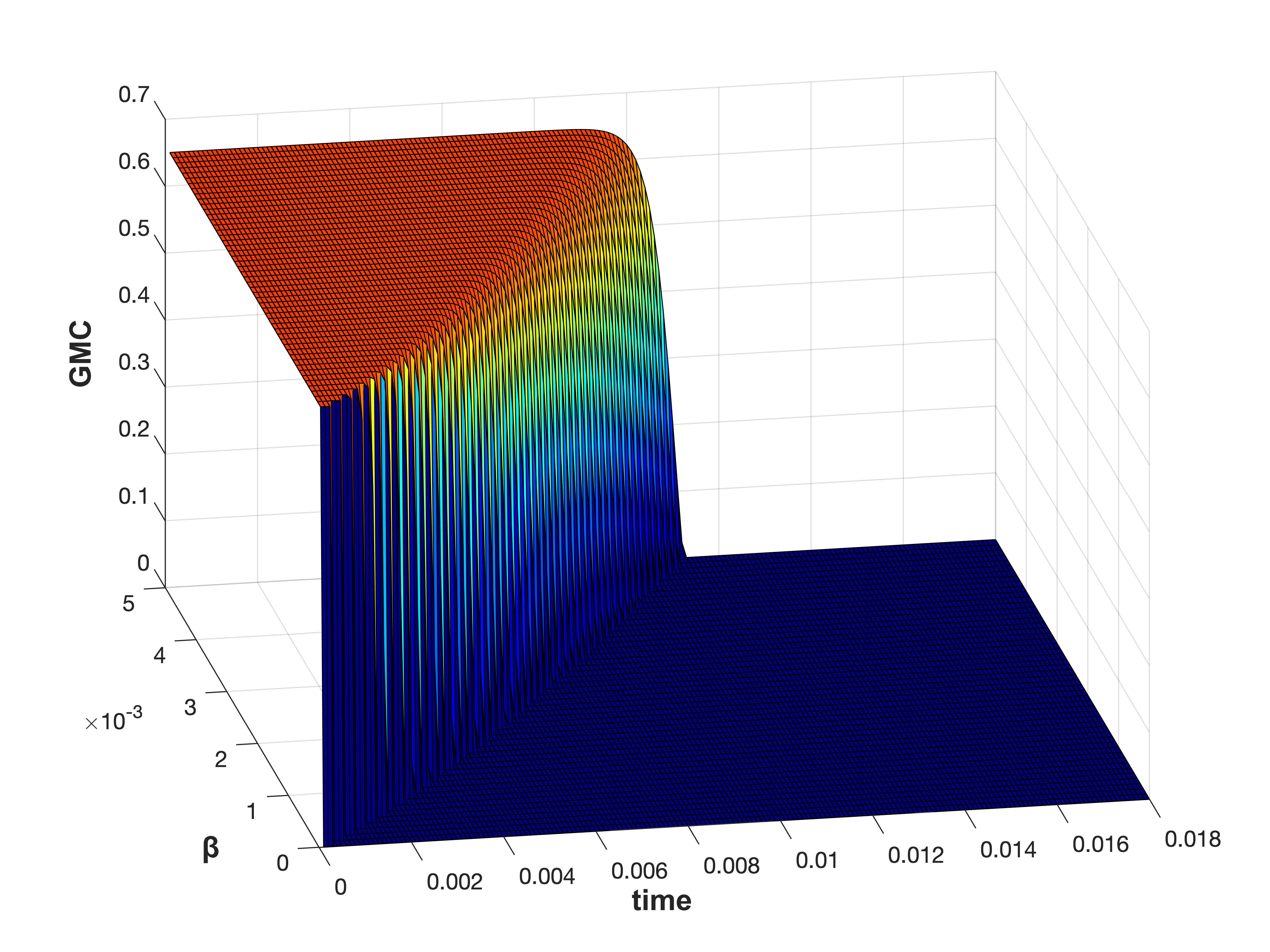}}
\hspace{\fill}
   \subfloat[\label{}]{%
      \includegraphics[trim=20 40 50 30,clip, width=0.3\textwidth]{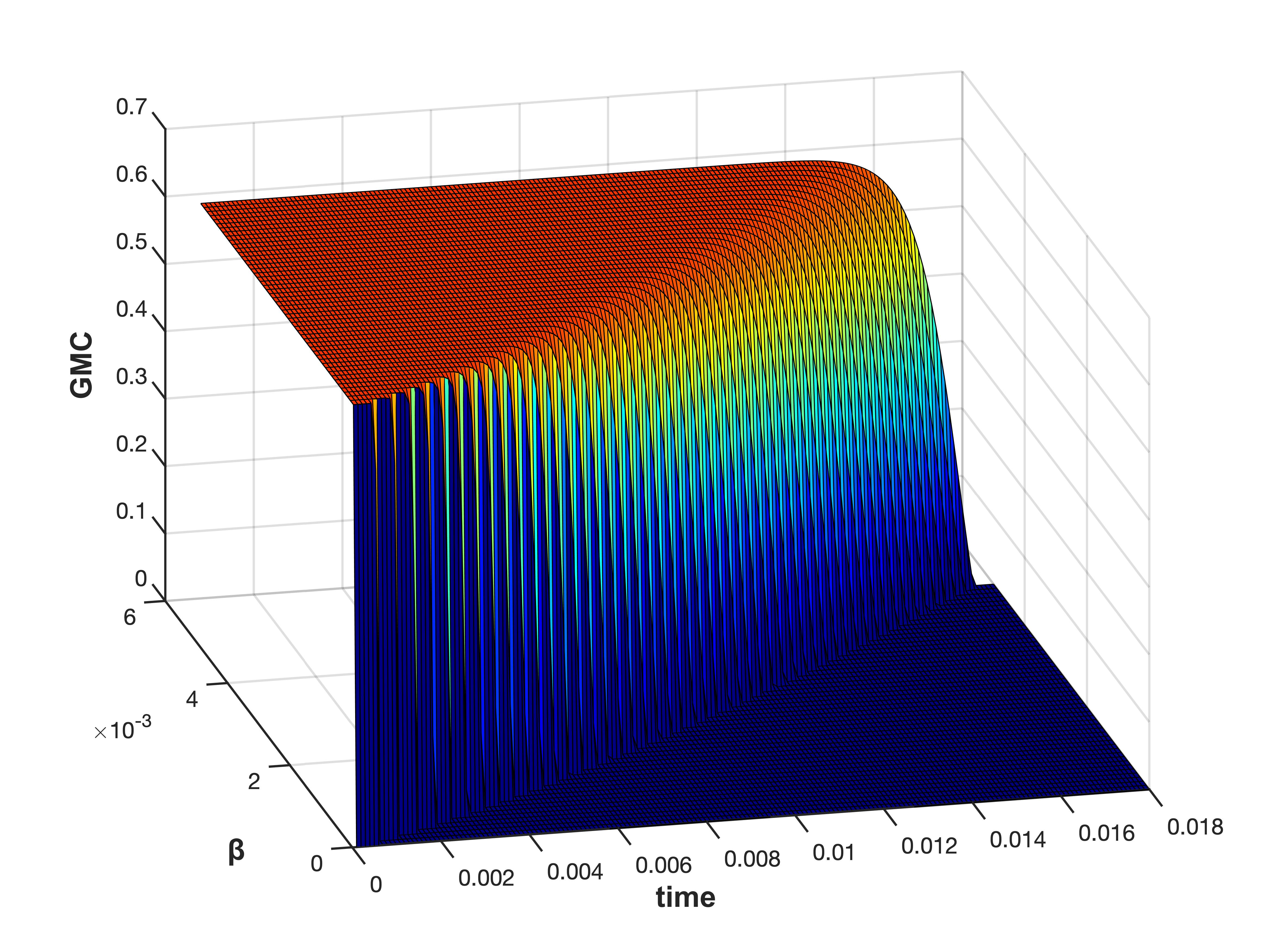}}\\
\caption{3D plot of Genuine multipartite concurrence (GMC) as a function of inverse temperature $\beta$ and time for Ohmic reservoir at a fixed mixing paramter $x = 0.8$, $\eta = 0.4$ and $\Omega^{2} = 36 $ along with $\beta = \beta_{A}$, $\beta_{B} = k_{1}\beta_{A}$, $\beta_{C} = k_{2} \beta_{A}$. (a) $k_{1} = 1.0$ and $k_{2} = 1.0$ (b) $k_{1} = 1.0$ and $k_{2} = 2.0$ } (c) $k_{1} = 4.0$ and $k_{2} = 16.0$\label{3d_plot_2}
\end{figure*}

From Fig. (\ref{GMC_ohmic_differnt_temperatures}) and Fig. (\ref{3d_plot_2}), we observe that the maximum value of genuine multipartite concurrence does not change with temperature, however, the preservation time is changed. In the Fig. (\ref{GMC_ohmic_differnt_temperatures}), we compare the different value of $k_{1}$ and $k_{2}$ with the case of $k_{1} = k_{2} = 1$ i.e., all three reservoirs are at equal temperature. We observe that there is a sharp increase in the preservation time as we increase the value of $k_{1}$ and $k_{2}$. Furthermore, larger the value of the product $k_{1}k_{2}$, the longer is the preservation time for a given value of $\beta$. Interestingly, the preservation time for a given $\beta$ does not increase linearly with the product $k_{1}k_{2}$. In fact, as one increases the value of $k_{1}k_{2}$, the increase in the preservation time for a given $\beta$ gets smaller and eventually saturates. Thus, in order to make correlations of the system of three qubits placed in independent reservoirs last for a significant amount of time, one has to increase the temperature gradient between the reservoirs and at the same time maximize the value of the product $k_{1}k_{2}$. Therefore, the presence of a temperature gradient plays a significant role in robustness of multipartite correlation in the three qubit systems, and  an increase in temperature gradient increases the preservation time which is in striking contrast to the case when all the reservoirs have same temperature and the correlation decays faster with increase in temperature.

\subsection{Evolution of tripartite negativity in W class Werner state}
 W state is a highly entangled tripartite state which belongs to a different i.e.,
 \begin{equation}\label{winitial}
     \rho(0) = (1-x)\frac{I}{8} + x|W\rangle\langle W|.
 \end{equation}
 As in the earlier cases, the diagonal elements will remain invariant and the time evolved off-diagonal elements are obtained as,
 \begin{equation}\label{den_mat_off}
    \begin{aligned}
        \rho_{(001)(010)}(t) &= \frac{x}{3}\operatorname{e}^{-(\Gamma_{B}(t) + \Gamma_{C}(t))}\operatorname{e}^{-i(E_{001}-E_{010})t},\\
        \rho_{(001)(100)}(t) &= \frac{x}{3}\operatorname{e}^{-(\Gamma_{A}(t) + \Gamma_{C}(t))}\operatorname{e}^{-i(E_{001}-E_{100})t},\\
        \rho_{(010)(100)}(t) &= \frac{x}{3}\operatorname{e}^{-(\Gamma_{A}(t) + \Gamma_{B}(t))}\operatorname{e}^{-i(E_{010}-E_{100})t}.
    \end{aligned} 
 \end{equation}

As described in the section \ref{prelim}, the tripartite negativity is obtained through taking geometric mean of the negativity corresponding to the three bipartitions.  The negativity corresponding to the bipartition $A|BC, B|AC$ and $C|AB$, are obtained as,
\begin{equation}
\begin{aligned}
    &\mathcal{N}_{A|BC} = 2\max\{0, \frac{x\operatorname{e}^{-\Gamma(t)}}{3}\sqrt{\operatorname{e}^{-2\Gamma_{C}(t)} + \operatorname{e}^{-2\Gamma(t)}} - \frac{1-x}{8}\},\\
    &\mathcal{N}_{B|AC} = 2\max\{0, \left(\frac{x^{2}\operatorname{e}^{-2(\Gamma(t) +\Gamma_{C}(t))}}{9} - X - \frac{x+3}{24}\right)\},\\
    &\mathcal{N}_{C|AB} = 2\max\{0, \frac{\sqrt{2}x\operatorname{e}^{-(\Gamma(t) + \Gamma_{C}(t))}}{3} - \frac{1-x}{8}\}.
\end{aligned}
\end{equation}
where, we have assumed $\Gamma(t) = \Gamma_{A}(t) =\Gamma_{B}(t) $, and 
\begin{equation}
    X = \sqrt{36m^{2}n^{2} + 4m^{2}x^{2} + 9n^{2} + \frac{n^{2}}{2} +\frac{x^{2}}{36}},
\end{equation}
with $m = x\exp(-(\Gamma_{C}(t) + \Gamma(t)))/3$ and $n = x\exp(-(2\Gamma(t)))/3$.
 Therefore, the tripartite negativity is obtained through,
 \begin{equation}
     \mathcal{N}(\rho) = \left(\mathcal{N}_{A|BC}\mathcal{N}_{B|AC}\mathcal{N}_{C|AB}\right)^{\frac{1}{3}}.
 \end{equation}

\begin{figure}[ht]
   \begin{subfigure}{0.5\textwidth}
\centering
\includegraphics[width=0.85\textwidth, height = 5cm]{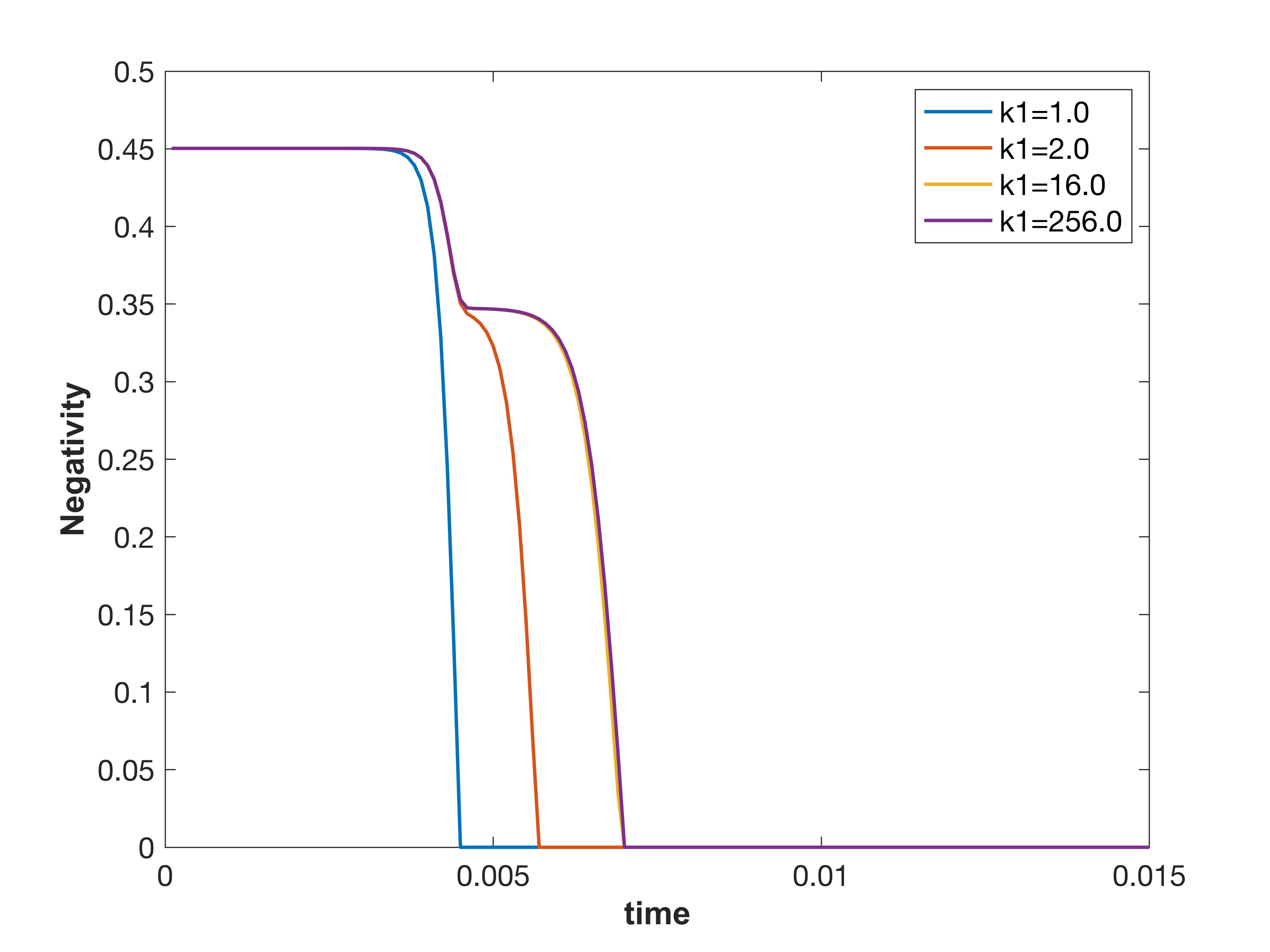} 
\caption{}
\end{subfigure}
\begin{subfigure}{0.5\textwidth}
\centering
\includegraphics[width=0.85\textwidth, height = 5cm]{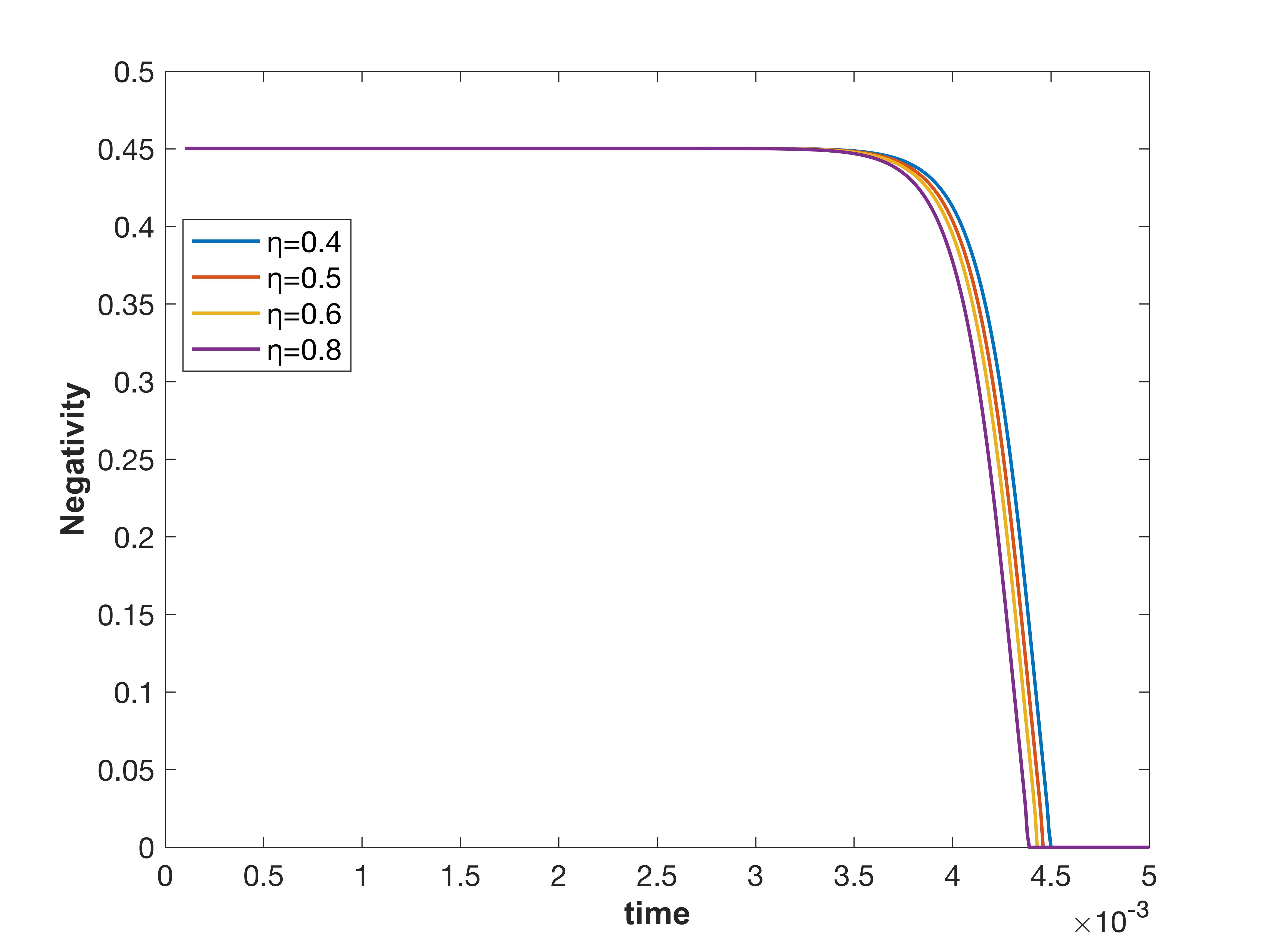}
\caption{}
\end{subfigure}
\caption{Negativity is plotted against time for ohmic spectral density for $\Omega^{2} = 36$, $x = 0.6$ and $\beta_{C} = k_{1}\beta_{A} = k_{1}\beta_{B}$ (a) for different values of $k_{1}$ at a fixed $\eta = 0.4$. With increase in $k_{1}$, preservation time increases. (b) for different values of $\eta$ at fixed value of $k_{1} = 1.0$.}\label{nega}
\end{figure}

Tripartite negativity is plotted for time evolved density matrix with initial tripartite state in Eq. (\ref{winitial}) in the Fig. (\ref{nega}) for various coupling constants, and for various temperature gradient between reservoirs.  
\begin{figure}[h!]
    \centering
    \includegraphics[width=0.5\textwidth]{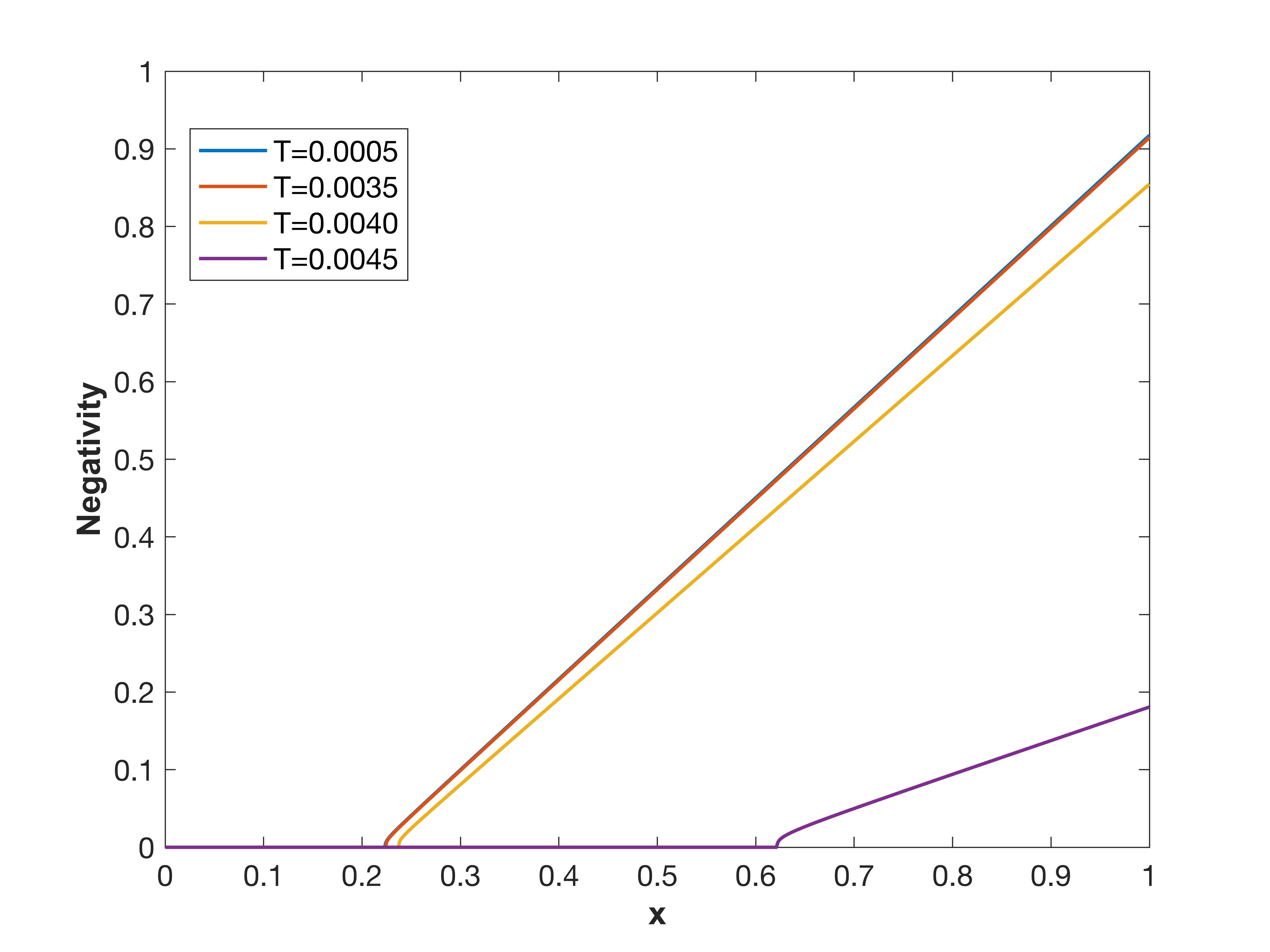}
    \caption{Negativity plotted against mixing parameter $x$ for ohmic spectral density with parameters $\Omega^{2} = 36$, $\eta = 0.4$ and for $\beta_{A} = \beta_{B} = \beta_{C} = 0.002$ at fixed time. }\label{N_vs_x}
\end{figure}
To do a comparative analysis of tripartite negativity with the decoherence in the state, we make use of a well defined coherence measure  $l_{1}$ norm coherence \cite{Plenio_PRL_2014}, which for a density matrix $\rho$ is defined through  the sum of the absolute value of the off diagonal elements of density matrix i.e.,
 \begin{equation}
     C_{l_{1}}(\rho) = \sum_{i\neq j}|\rho_{ij}|.
 \end{equation}
 
 \begin{figure}[ht!]
   \begin{subfigure}{0.5\textwidth}
\centering
\includegraphics[width=0.85\textwidth, height = 5cm]{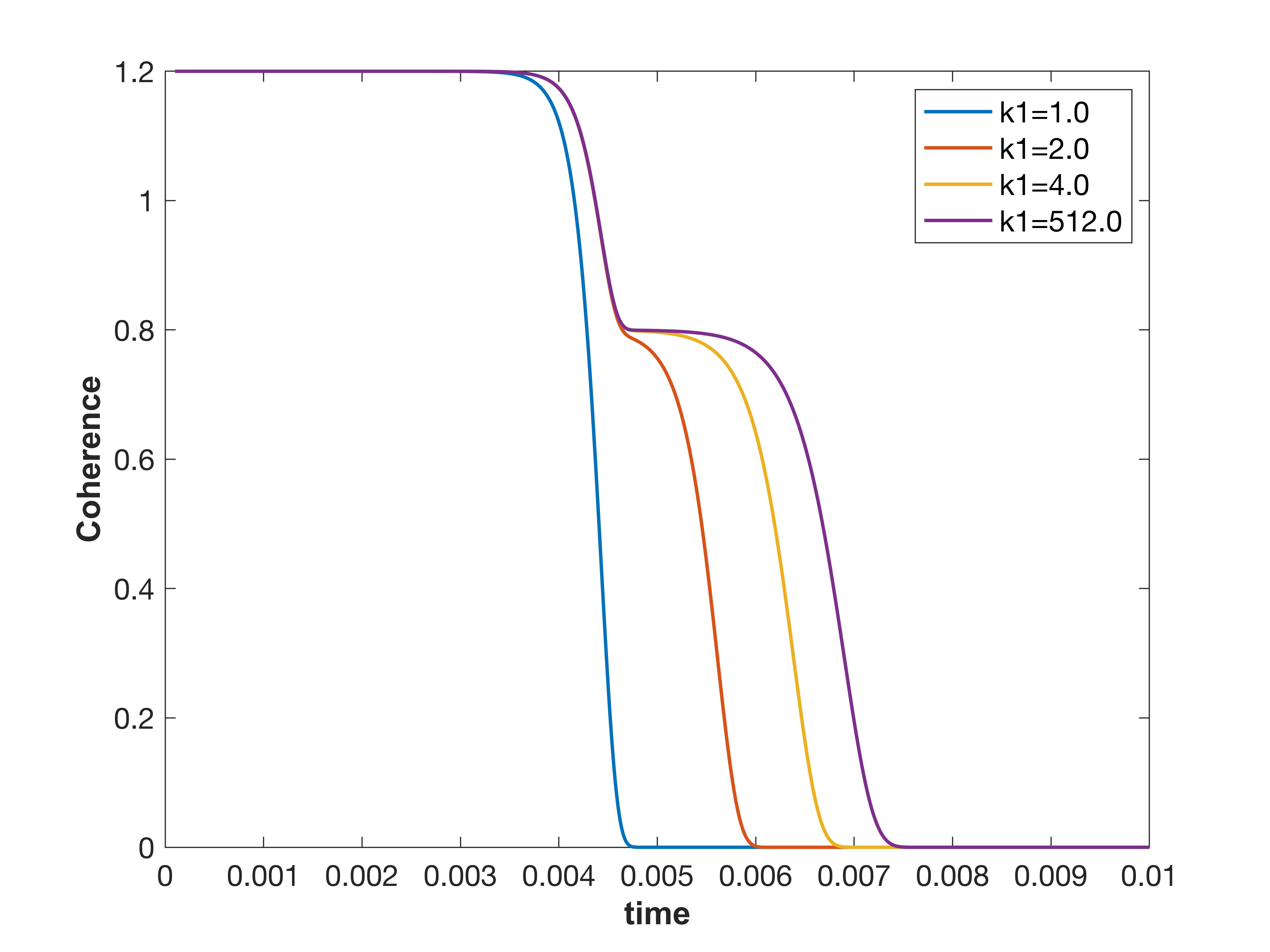} 
\caption{}
\end{subfigure}
\begin{subfigure}{0.5\textwidth}
\centering
\includegraphics[width=0.85\textwidth, height = 5cm]{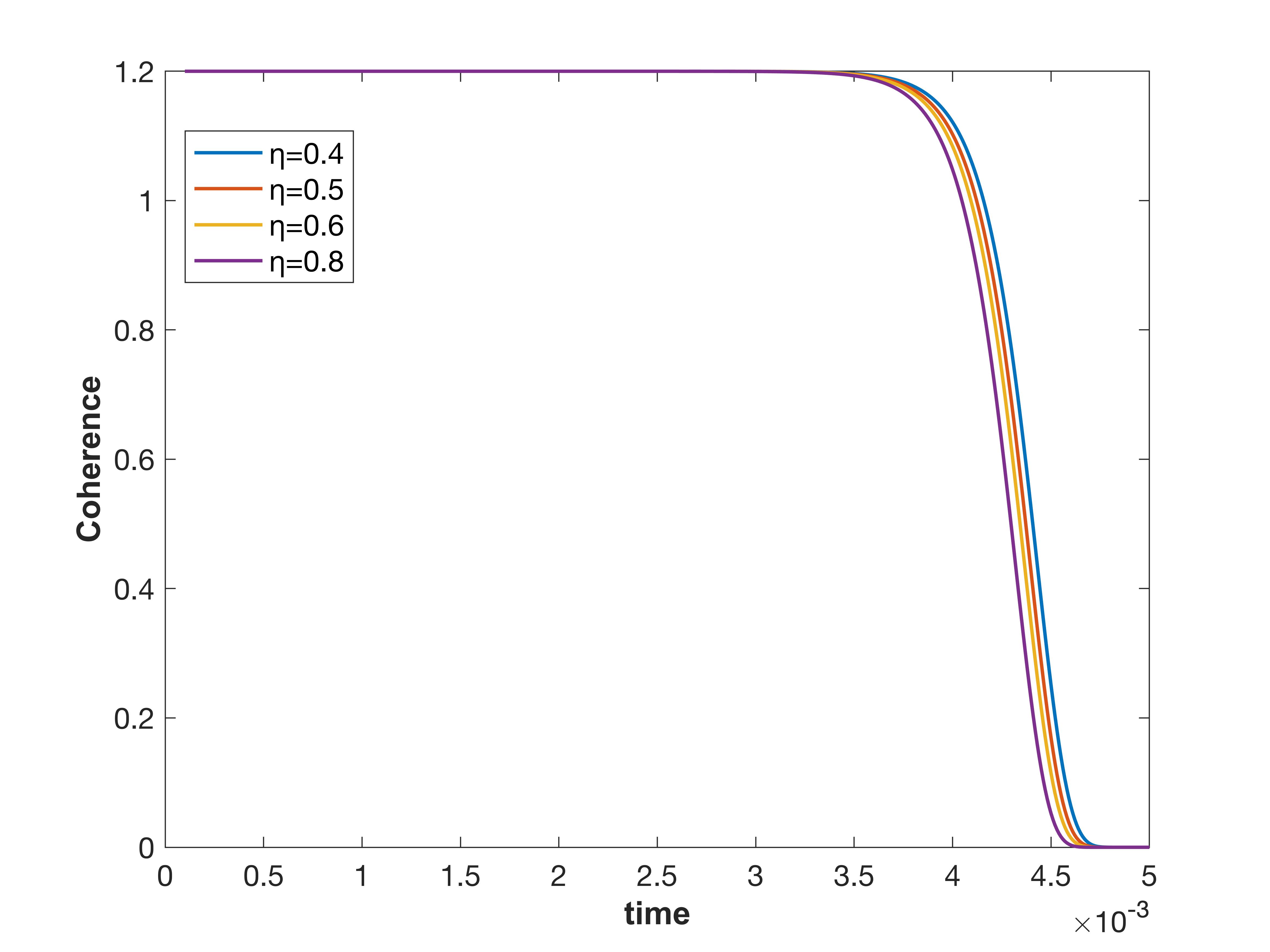}
\caption{}
\end{subfigure}
\caption{Coherence is plotted against time for ohmic spectral density for parameters $\Omega^{2} = 36$, $x = 0.6$ and $\beta_{C} = k_{1}\beta_{A} = k_{1}\beta_{B}$ (a) for different values of $k_{1}$ at a fixed $\eta = 0.4$. (b) for different values of $\eta$ at fixed value of $k_{1} = 1.0$.}\label{coher}
\end{figure}

\begin{figure}[h!]
    \centering
    \includegraphics[width=0.5\textwidth]{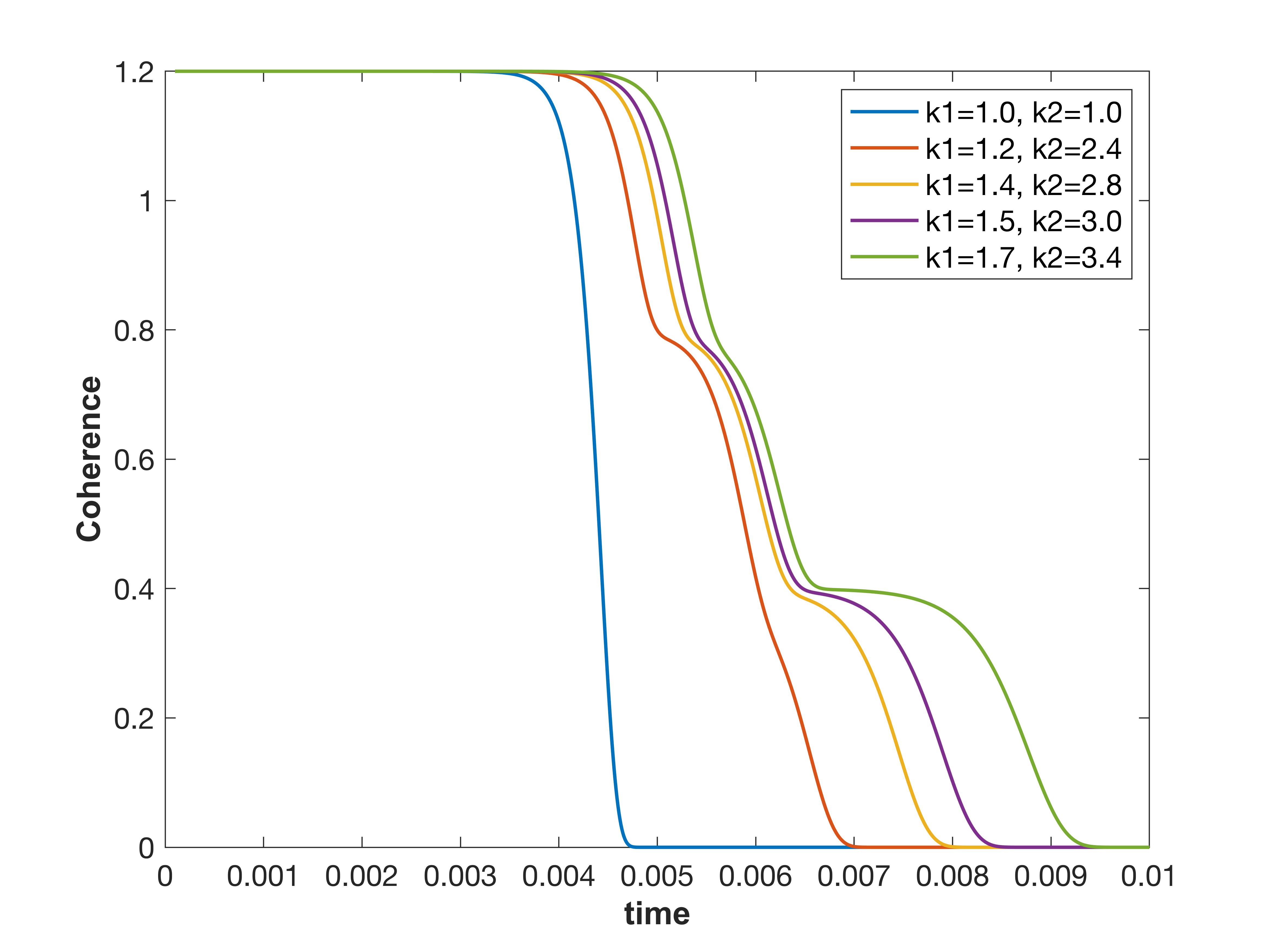}
    \caption{Coherence as measured through $l_{1}$ norm plotted against time for ohmic spectral density with parameters $\Omega^{2} = 36$, $x = 0.6$, $\beta_{A} = 0.003$ and $\beta_{B} = k_{1}\beta_{A}$ and $\beta_{C} =  k_{2}\beta_{A}$.}\label{coh_diff_k}
\end{figure}

 Since $\Gamma_{X}(t)$ is a real, positive and increasing function of time, the absolute value of the off diagonal elements of density matrix (\ref{den_mat_off})  decays exponentially. Therefore,  the coherence remains finite and becomes zero asymptotically. In contrast, the tripartite negativity reaches zero after a finite time, hence the state experiences sudden death of tripartite negativity. From Fig. (\ref{nega}) and (\ref{coher}), it is observed that the coherence and negativity of the tripartite system remains constant for the time interval $0<t<T_{c}$ and then decay rapidly for $t > T_{c}$, where $T_{c}$ is the characteristic time. Moreover, greater the temperature gradient between the reservoirs the larger is the characteristic time. Interestingly, as we increase the value of $k_{1}$, the increase in the characteristic time for a given value of $\beta$ becomes minute for the case of tripartite negativity as compared to coherence,  however the characteristic time eventually saturates for both the measures. From Figure (\ref{N_vs_x}) it is observed that the value of mixing parameter $x$ at which negativity becomes zero, increases with time $t$. This increase is very slows at first ($0.0005<t<0.0035$) but becomes sharp later ($t > 0.004$). This behaviour is in contrast to genuine multipartite concurrence (GMC) which is zero for mixing parameter $x < 3/7$ at all values of time. Furthermore, for large temperature gradients, we observe that after a steep decay in the coherence, there is a small interval for which the coherence shows robustness in multiple regions (see Fig. (\ref{coh_diff_k})), and  consequently a similar robustness is shown by the tripartite negativity (see Fig. (\ref{nega})). Therefore, for a large temperature gradient, in addition to an increase in the preservation duration of multipartite correlations, the state also experiences the robustness of correlation in multiple intervals of time.

\section{Conclusion}\label{conclusion}
In conclusion, we have investigated the dynamics of tripartite correlation in three qubit systems, with the qubits present in independent thermal reservoirs. We model the reservoirs as set of infinite quantum harmonic oscillators, and with a bilinear form of non-dissipative interaction Hamiltonian, derived an exact expression for the time evolved density matrix. We studied the genuine multipartite concurrence for GHZ class Werner state, and showed that the correlation shows robustness up to a characteristic time and then decays steeply. Further, preservation time for GMC is shown to increase with an increase in temperature gradient, however, it is shown to saturate for very high temperature gradients. We further studied the dynamics of $l_{1}$ norm coherence and tripartite negativity for W class Werner state, and showed that for the case with vanishing temperature gradient between reservoirs, a similar behaviour of robustness up to a characteristic time and then sharp decay is observed. Interestingly, when there is a temperature gradient, we showed that one observes multiple regions where the coherence and negativity shows robustness against environmental decoherence. Therefore, with an increase in temperature gradient, in addition to increase in preservation time, we observed multiple region with freezing behavior of tripartite correlation.  
\section*{Acknowledgement}
AKR and PKP acknowledge the support from DST, India
through Grant No. DST/ICPS/QuST/Theme-1/2019/2020-
21/01. 
\begin{spacing}{1.1}
\bibliographystyle{spmpsci_unsort} 
\bibliography{sample}  
\end{spacing}

\end{document}